\documentclass[twocolumn]{aastex6}
\usepackage{rotate}
\usepackage{amsmath}
\usepackage{times}
\usepackage{booktabs}
\usepackage{subfigure}
\usepackage{textcomp}
\usepackage{multirow}
\usepackage{gensymb}
\usepackage{graphicx}	

\def\xmm {\emph{XMM--Newton}}
\def\nustar {\emph{NuSTAR}}
\def\chandra {\emph{Chandra}}

\def\flux {\mbox{erg\,cm$^{-2}$\,s$^{-1}$}}
\def\lum {\mbox{erg\,s$^{-1}$}}
\def\src {\mbox{M51\,ULX-7}}

\begin{document}
\accepted{15-Apr-2020}
\title{Discovery of a 2.8\,\MakeLowercase{s} pulsar in a 2\,\MakeLowercase{d}-orbit High-Mass X-ray Binary powering the Ultraluminous X-ray source ULX-7 in M51}
\shorttitle{Discovery of a 2.8\,\MakeLowercase{s} PULX in M51}
\author{G.\,A. Rodr\'iguez Castillo\altaffilmark{1}, G.\,L.\,Israel\altaffilmark{1}, A.\,Belfiore\altaffilmark{2} , F.\,Bernardini\altaffilmark{1,3,4}, P.\,Esposito\altaffilmark{5,2}, F.\,Pintore\altaffilmark{2}, A. De Luca\altaffilmark{2,6},  A.\,Papitto\altaffilmark{1}, L.\,Stella\altaffilmark{1}, A.\,Tiengo\altaffilmark{5,2,6}, L.\,Zampieri\altaffilmark{7}, M.\,Bachetti\altaffilmark{8}, M.\,Brightman\altaffilmark{9}, P.\,Casella\altaffilmark{1}, D.\,D'Agostino\altaffilmark{10}, S.\,Dall'Osso\altaffilmark{11}, H.\,P.\,Earnshaw\altaffilmark{9}, F.\,F\"urst\altaffilmark{12}, F.\,Haberl\altaffilmark{13}, F.\,A.\,Harrison\altaffilmark{9}, M.\,Mapelli\altaffilmark{7,14,15,16}, M.\,Marelli\altaffilmark{2},  M.\,Middleton\altaffilmark{17}, 
C.\,Pinto\altaffilmark{18}, T.\,P.\,Roberts\altaffilmark{19}, R.\,Salvaterra\altaffilmark{2}, R.\,Turolla\altaffilmark{14,20}, D.\,J.\,Walton\altaffilmark{21} and  A.\,Wolter\altaffilmark{22}}

\shortauthors{Rodr\'iguez Castillo et al.}

\altaffiltext{1}{INAF--Osservatorio Astronomico di Roma, via Frascati 33, 00078 Monteporzio Catone, Italy}
\altaffiltext{2}{INAF--Istituto di Astrofisica Spaziale e Fisica Cosmica di Milano, via A. Corti 12, 20133 Milano, Italy}
\altaffiltext{3}{INAF--Osservatorio Astronomico di Capodimonte, salita Moiariello 16, 80131 Napoli, Italy}
\altaffiltext{4}{New York University Abu Dhabi, Saadiyat Island, PO Box 129188 Abu Dhabi, United Arab Emirates}
\altaffiltext{5}{Scuola Universitaria Superiore IUSS Pavia, piazza della Vittoria 15, 27100 Pavia, Italy}
\altaffiltext{6}{Istituto Nazionale di Fisica Nucleare (INFN), Sezione di Pavia, via A. Bassi 6, 27100 Pavia, Italy}
\altaffiltext{7}{INAF--Osservatorio Astronomico di Padova, vicolo dell'Osservatorio 5, 35122 Padova, Italy}
\altaffiltext{8}{INAF--Osservatorio Astronomico di Cagliari, via della Scienza 5, 09047 Selargius (CA), Italy}
\altaffiltext{9}{Cahill Center for Astronomy and Astrophysics, California Institute of Technology, 1216 East California Boulevard, Pasadena, CA 91125, USA}
\altaffiltext{10}{Istituto di Matematica Applicata e Tecnologie Informatiche (IMATI) `E. Magenes', CNR, via de Marini 6, 16149 Genova, Italy}
\altaffiltext{11}{Department of Physics and Astronomy, Stony Brook University, Stony Brook 11579 NY, USA}
\altaffiltext{12}{European Space Astronomy Centre (ESAC), ESA, Camino Bajo del Castillo s/n, Villanueva de la Ca\~nada, 28692 Madrid, Spain}
\altaffiltext{13}{Max-Planck-Institut f{\"u}r extraterrestrische Physik, Gie{\ss}enbachstra{\ss}e 1, 85748 Garching, Germany}
\altaffiltext{14}{Dipartimento di Fisica e Astronomia `Galileo Galilei', Universit\`a di Padova, via F. Marzolo 8, 35131 Padova, Italy}
\altaffiltext{15}{Institut f\"ur Astro- und Teilchenphysik, Universit\"at Innsbruck, Technikerstrasse 25/8, 6020, Innsbruck, Austria}
\altaffiltext{16}{Istituto Nazionale di Fisica Nucleare (INFN), Sezione di Padova, via F. Marzolo 8, 35131 Padova, Italy}
\altaffiltext{17}{Department of Physics and Astronomy, University of Southampton, Highfield, Southampton SO17 1BJ, UK}
\altaffiltext{18}{European Space Research and Technology Centre (ESTEC), ESA, Keplerlaan 1, 2201\,AZ Noordwijk, The Netherlands}
\altaffiltext{19}{Centre for Extragalactic Astronomy, Department of Physics, Durham University, South Road, Durham DH1 3LE, UK}
\altaffiltext{20}{Mullard Space Science Laboratory, University College London, Holmbury St. Mary, Dorking, Surrey RH5 6NT, UK}
\altaffiltext{21}{Institute of Astronomy, Science Operations Department, University of Cambridge, Madingley Road, Cambridge CB3 0HA, UK}
\altaffiltext{22}{Osservatorio Astronomico di Brera, INAF, via Brera 28, 20121 Milano, Italy}

\begin{abstract}
We discovered 2.8\,s pulsations in the X-ray emission of the ultraluminous X-ray source (ULX) \src\ within the UNSEeN project, which was designed to hunt for new pulsating ULXs (PULXs) with \xmm. The pulse shape is sinusoidal and large variations of its amplitude were observed even within single exposures (pulsed fraction from less than 5\% to 20\%).
\src\ is a variable source, generally observed at an X-ray luminosity between $10^{39}$ and  $10^{40}$\,\lum, located in the outskirts of the spiral galaxy M51a at a distance of 8.6\,Mpc. 
According to our analysis, the X-ray pulsar orbits in a 2-d binary with  a projected semi-major axis $a_\mathrm{X} \sin i \simeq$ 28\,lt-s. For a neutron star (NS) of 1.4\,$M_{\odot}$, this implies a lower limit on the companion mass of 8\,$M_{\odot}$, placing the system hosting \src\, in the high-mass X-ray binary class. 
The barycentric pulse period decreased by $\simeq$0.4\,ms in the 31\,d spanned by our  May--June 2018 observations, corresponding to a spin-up rate $\dot{P} \simeq -1.5\times10^{-10}\text{ s s}^{-1}$. In an archival 2005 \xmm\ exposure, we measured a spin period of $\sim$3.3\,s, indicating a secular spin-up of $\dot{P}_{\mathrm{sec}}\simeq -10^{-9}\text{ s s}^{-1}$, a value in the range of other known PULXs. Our findings suggest that the system consists of a massive donor, possibly an OB giant or supergiant, and a  moderately magnetic (dipole field component in the range $10^{12}$\, G $\lesssim B_{\mathrm{dip}}\lesssim 10^{13}$\,G) accreting NS with weakly beamed emission ($1/12\lesssim b\lesssim1/4$).  
\end{abstract}

\keywords{accretion, accretion disks --- galaxies: individual: M51a --- pulsars: individual (\src, aka CXOM51\,J133001.0+471344) --- stars: neutron --- X-rays: binaries}

\section{Introduction}

 Ultraluminous X-ray sources are off-nucleus objects detected in nearby galaxies with  X-ray luminosities in excess of $10^{39}$~\lum, the Eddington luminosity  ($L_{\mathrm{Edd}}$) of a 10\,$M_\odot$ object \citep[e.g.][for a review]{kaaret17}. Under the assumption of a stationary, spherically symmetric accretion flow, 
$L_{\mathrm{Edd}}$ sets an upper limit to the accretion luminosity 
that a compact object can steadily produce, since 
for higher values the accretion flow would be halted by radiation forces. For a compact object accreting fully ionized hydrogen the above limit can be written as $L_{\mathrm{Edd}}=4\pi cG M  m_{\mathrm{p}}/\sigma_{\mathrm{T}}\simeq1.3\times10^{38} (M/M_{\odot})$~\lum, 
where $\sigma_{\mathrm{T}}$ is the Thomson scattering cross section, $m_{\mathrm{p}}$ 
is the proton mass, and $M$ is the mass of the compact object. 
Since early discoveries in the `70s with the \emph{Einstein} mission \citep{long83,fabbiano92}, the high luminosity of ULXs has been interpreted in terms of accretion 
at or  above the Eddington limit onto BHs of stellar origin 
($<$80--100\,$M_\odot$; e.g. \citealt{stobbart06,roberts07,zampieri09,feng11}), or sub-Eddington accretion onto intermediate-mass BHs (IMBH, $10^3$ -- $10^5$\,$M_\odot$; e.g. \citealt{colbert99,sutton12}).

The recent discovery of coherent pulsations with periods between 
0.4 and 40\,s  in the X-ray light curves of four distinct ULXs with 
luminosities in the $10^{39}$ -- $10^{41}$\,\lum\ range, unambiguously 
associate these ULXs with accreting NSs, i.e. compact objects with 
mass of only $\sim$1--2 $M_{\odot}$ \citep{bachetti14,furst16,israel17,ipe17,carpano18}. 
Recently a new candidate PULXs, NGC\,1313\,X-2, has been reported based on weak and transient pulsations at a period of $\sim$1.5\,s \citep{sath19}.
These X-ray pulsars demonstrate that accreting NSs can attain extreme luminosities, above 500 
times  $L_{\mathrm{Edd}}$, which are difficult to interpret in the context of 
standard accretion models for NS X-ray binaries. A 
significantly super-Eddington
luminosity can be achieved if the magnetic field of the NS is very high,
as a result of a marked reduction of the opacities for extraordinary photons: in particular, a luminosity of 
$\sim$500\,$L_{\mathrm{Edd}}$ can be attained for a field strength 
$>$\,$10^{15}$\,G \citep{mushtukov15,dallosso15}, which is expected in magnetars \citep[see e.g.][for a review]{turolla15}. 
Rotation of the NS and its magnetosphere drags matter at the
magnetospheric boundary; if rotation is fast enough  the centrifugal force exceeds 
the gravitational force locally and inhibition of accretion on the NS surface results from the 
so-called propeller mechanism \citep{illarionov75,stella86}. Owing to their relatively fast spin period, invoking very strong magnetic fields would imply that the propeller mechanism operates in PULXs; this can be mitigated by assuming that the emission is beamed. 
In the most luminous PULX, namely NGC\,5907 ULX, a beaming factor of 1/100 would 
be needed \citep{israel17}. 

Several scenarios have been proposed to account 
for the PULX properties. The presence of a strong multipolar magnetic 
field ($\sim$10$^{14}$\,G) close to the surface of the NSs coupled with a modest 
degree of beaming appears as a reasonable way out of the problem \citep{israel17,chashkina17}. In this scenario, a standard magnetic dipole field of $\sim$10$^{12}-$10$^{13}$\,G dominates, at large distances, over the multipolar component, the effect of which are limited to the region close to the surface and to the accretion column base.  However, this scenario does not necessarily invoke the presence of magnetars, 
since magnetars are not merely young NSs with a high dipolar field and/or with a multipolar component; they are also characterized by specific bursting and flaring activity (see e.g. \citealt{eri18}).
Alternatively, standard magnetic fields (in the $10^{11}-10^{13}$~G range,  without any significant multipolar component) are envisaged by models 
in which the disk is fed at a super-Eddington rate, the excess supply is ejected away, and the emission is highly beamed in a geometrically thick inner disk funnel \citep[see e.g.][]{king16,king17,Pintore17,koliopanos17,walton18b,king19}.
We note that two more extragalactic transient pulsators share striking similarities with the above group, i.e. super-Eddington luminosities and/or large first derivative of the spin period, namely
XMMU\,J031747.5--663010 in NGC1313 ($\sim$766\,s; \citealt{trudolyubov08}) 
and  CXOU\,J073709.1+653544 in NGC2403 ($\sim$18\,s; \citealt{trudolyubov07}). 
We suggest here for the first time that they could be two other PULXs that went unnoticed so far.

The discovery of  PULXs calls into question the nature ULXs, many of which have been classified as accreting black holes due to their high luminosity: 
in fact an unknown, but possibly large fraction of ULXs may host an accreting NS rather than a BH \citep{wiktorowicz17,middleton17,middleton17e}. Therefore,  assessing the nature of the compact objects hosted in ULXs is a key point in understanding the ULX population. In general, the unambiguous identification of the presence of a NS is achieved in X-rays by means of the detection of coherent periodic signals reflecting the spin period of the NS. However, given the small average pulsed 
fractions (PFs)\footnote{Throughout this work, we define the PF as the semi-amplitude of a sinusoidal fit to the pulse profile divided by the source average count rate.} of the flux observed in most PULXs (5--15\% range), sensitive searches for pulsations require a 
large number of counts ($\sim$10,000 or higher; see below). For the large majority of the ULXs observed with \xmm, the number of counts collected is by far too low for the detection of pulsations with PFs as small as those observed so far for the fastest-spinning PULXs. Sufficient statistics are currently available only for about $\sim$15 ULXs, out of about 300 known observed with \xmm\ \citep{earnshaw18}; and remarkably, among all the ULXs with good enough statistics data sets $\sim$25\% of them are proven to be NSs.


To increase the number of ULXs for which a sensitive search for pulsations
can be carried out and to provide a first constraint on the incidence of NSs in ULXs, 
we observed for long exposures ($\sim100$\,ks) with \xmm\ the fields of eight nearby galaxies (\mbox{$3\,\mathrm{Mpc}\leq d\leq30$\,Mpc}) hosting a considerable number of ULXs
(\xmm\ Large Project UNSEeN; Ultraluminous Neutron Star Extragalactic
populatioN).     
Among these galaxies is Messier 51a (M51a), also known as NGC 5194 or the Whirlpool Galaxy, 
a face-on
spiral interacting with the dwarf galaxy M51b (NGC\,5195). It is located at a
distance of $(8.58\pm0.10)$\,Mpc \citep{mcquinn16} and hosts a large 
number of X-ray sources, including nine ULXs \citep{terashima04,brightman18}.
Based on optical studies, M51a was classified as a Seyfert\,2 galaxy \citep{stauffer82}.

\src\ (also known as NGC\,5194 X-7, \citealt{roberts00}; CXOM51\,J133001.0+47134, \citealt{terashima04}; NGC 5194/5 ULX7 \citealt{liu05}) was first
detected with the {\it Einstein Observatory} 
observations at a luminosity above
$10^{39}$\,\lum\ \citep{palumbo85} and located at an offset of about 2.3\,arcmin from
the central AGN, on the outskirts of a young open cluster on a spiral
arm of M51.  Deep observations with \chandra\ showed pronounced
variability ($\Delta L_{\mathrm{X}}/L_{\mathrm{X}} \geq 10$) and the presence of a
$\sim$7,620\,s period modulation (at high $L_{\mathrm{X}}$;   
\citealt{liu02}). A flux modulation was also observed in an 
\xmm\ exposure, though at a significantly different period of 
$\sim$5,900\,s \citep{dewangan05}. The variation in period strongly argued 
against an orbital origin and suggested the presence of some kind 
of quasi periodic oscillations (QPOs). The \xmm\ spectral 
properties and the fact that the source resides near a young 
massive star cluster with age $T\sim12$\,Myr \citep{abolmasov07}  
suggested instead that \src\ is a high mass X-ray binary (HMXB). More 
recently, based on a multi-wavelength study (from radio to hard 
X-rays), it was proposed that the source might be an IMBH accreting in the hard state; 
however an accreting NS could not be excluded \citep{earnshaw16}.

In Section\,\ref{sec:analysis} we report on the discovery of coherent pulsations in the X-ray flux of \src\ at a period of about 2.8\,s with a highly variable amplitude, unambiguously making this source a new member of the rapidly growing class of PULXs. Refined timing analysis allows us to infer an orbital period of about 2\,d and a lower limit to the companion star mass of about $8\,M_{\odot}$. Furthermore, the analysis of \xmm\ archival data makes it possible to infer the secular first period derivative of this new PULX. In Section\,\ref{counterparts} we make use of the most updated X-ray position of \src\ in order to further constrain the optical properties and the nature of its possible counterpart. 
Finally, in Section\,\ref{sec:discussion} we put the inferred properties of \src\ in the more general context of the proposed accretion models and of the possible nature of the binary system.

\section{\emph{XMM--Newton} data}
\label{sec:analysis}

\subsection{Observations and data reduction}\label{data_reduction}
Within the \xmm\ Large Program UNSEeN, we obtained a 78\,ks-long observation of the M51 galaxy followed by three more pointings (one 98\,ks and two 63\,ks-long) carried out about a month apart as part of the  Discretionary time of the Project Scientist (DPS).
Before our campaign, the galaxy had already been observed by \xmm\ on six occasions with shorter exposure times (see Table\,\ref{tab:detections}). In the archival observations,
the EPIC pn and MOS cameras \citep{struder01,turner01} were operated in
various modes, with different time resolutions (MOS), sizes of the field of 
view (MOS), and position angles (the UNSEeN field of view was set 
to avoid targeted sources to fall in CCD gaps). 
The Science Analysis Software (SAS) v.17.0.0 was used to process the raw
observation data files. Intervals of time with anomalously high particle
background were filtered out. 

For the timing analysis, photon event lists of each source were extracted from circular 
regions  with a radius aimed at minimizing the spurious contribution of nearby objects and diffuse emission
in the crowded field of M51. The background was 
estimated from a nearby, source-free circular region with the same radius.
For the timing analysis we mainly used the pn data (with a time resolution of 73\,ms), but also data acquired by the MOSs were used when the cameras were operated in Small Window mode  (time resolution of 300\,ms, that is, only in the three DPS observations). 
Times of arrival (ToAs) of the photons were shifted to the barycentre
of the Solar system with the SAS task BARYCEN (the \chandra\ position from \citealt{kuntz16} was used; $\rm RA=13^h30^m01\fs02$, $\rm Dec.=47\arcdeg 13\arcmin 43\farcs8$; J2000).

To study the spectra of \src, both EPIC pn and MOS data were used. 
Some soft diffuse emission from the host galaxy surrounds the point source. For this reason, we were particularly careful in selecting the source region size and in the choice of the region to evaluate the background. We finally settled for a circular region of 35$''$ for the source events and for an annular region of inner and outer radii of 50$''$ and 70$''$ centered on the source position for the background. Some faint point-like sources lay inside the background region and were excluded from the event selection. We checked that different choices for the region size or the background did not impact the spectral results, in particular in the observation that caught the ULX at the lowest flux (see Sect.\,\ref{section:spectra}).  
No significant spectral discrepancies were observed between the pn and 
MOS spectra inside each observation, therefore we combined the data using the SAS tool 
\mbox{{\sc epicspeccombine}}. We checked that the combined spectrum was consistent with the single pn and MOS spectra. The source photons were grouped to a minimum 
of 25 counts per spectral bin and spectra rebinned to preserve the intrinsic 
spectral resolution using the SAS tool \mbox{{\sc specgroup}}. 

\begin{figure*}[thb]
\centering
\resizebox{\hsize}{!}{\includegraphics{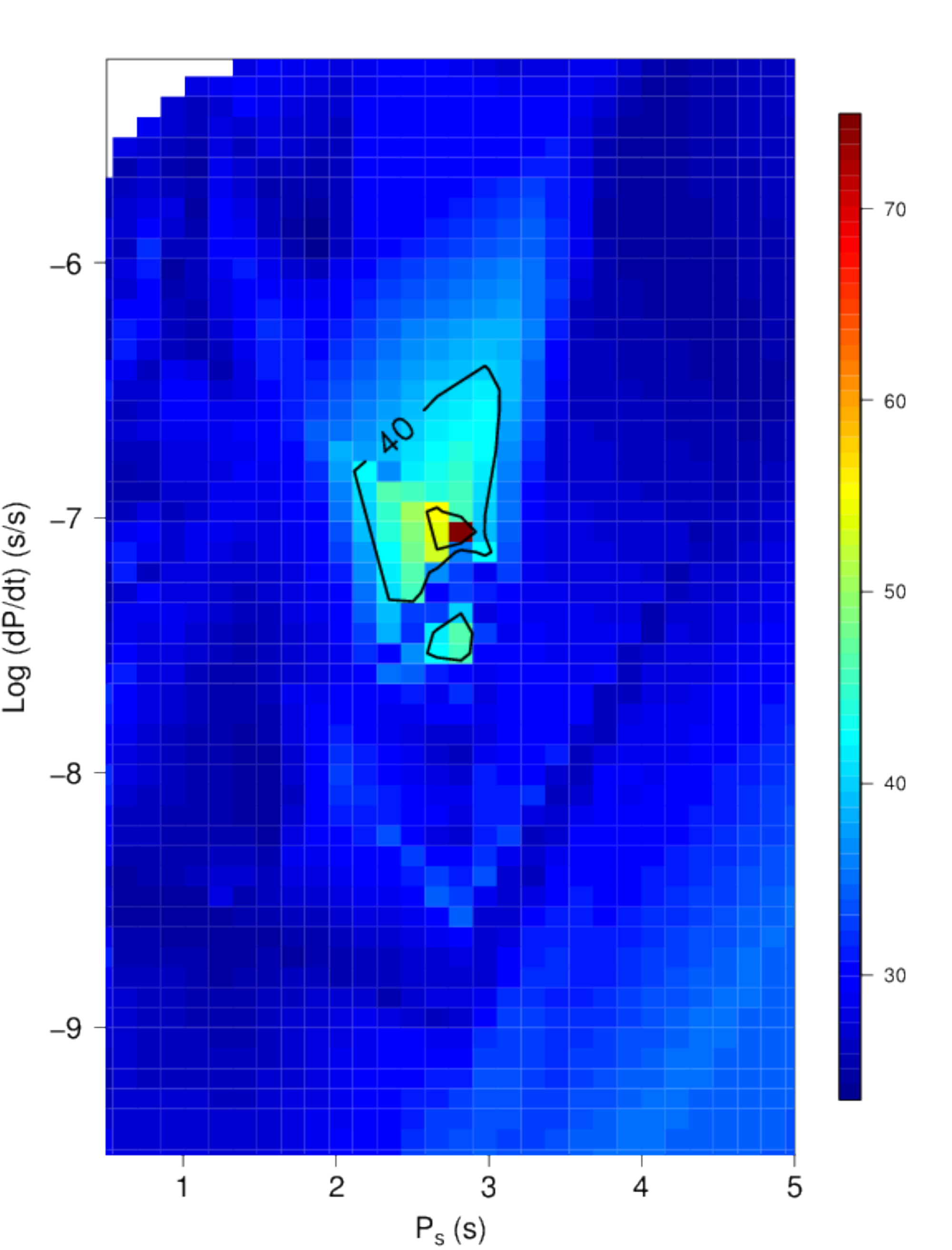}\includegraphics{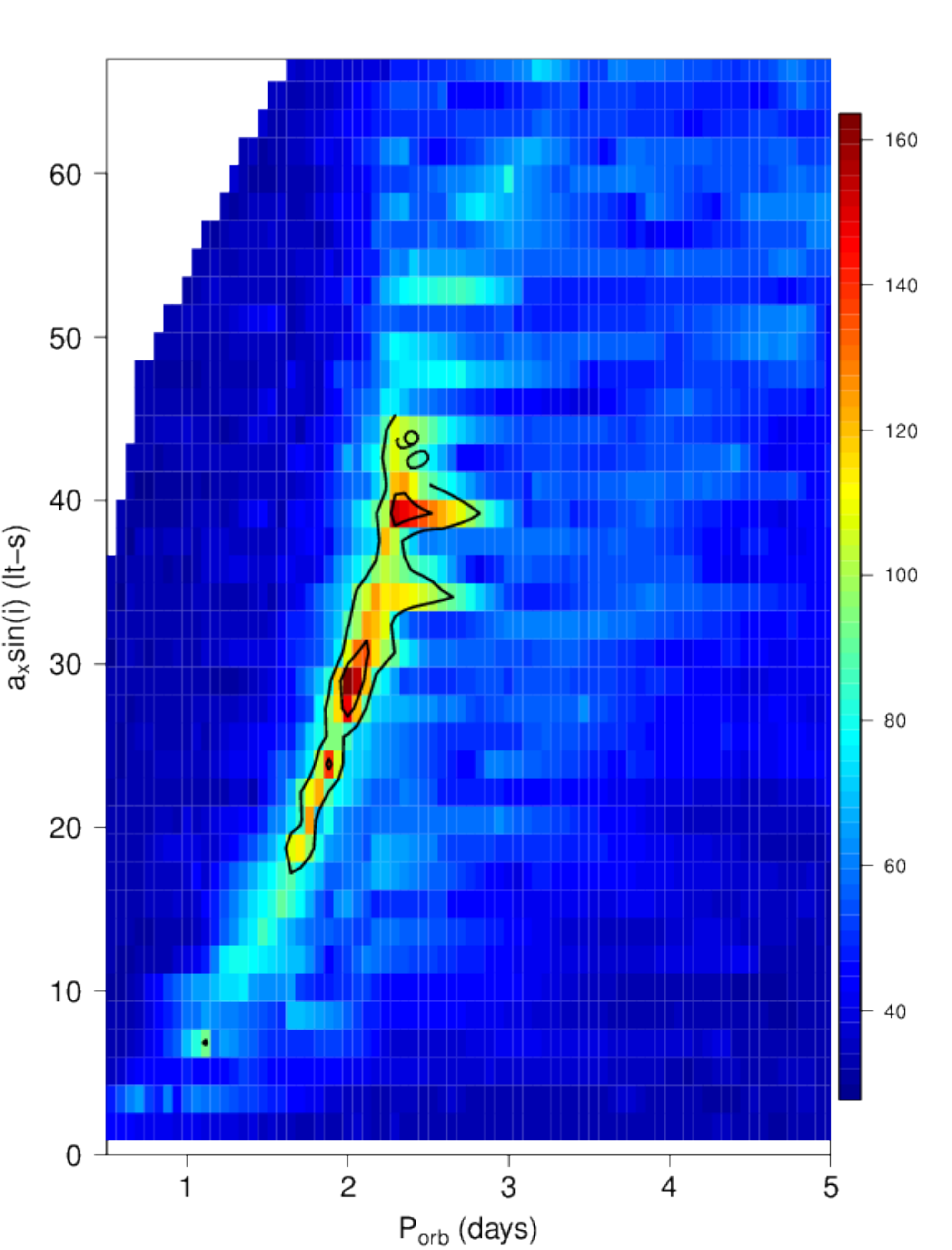}}
\caption{Left panel: PASTA discovery plot for the 2.8\,s-period signal in \src\ (observation 0824450901) where each point in the plane corresponds to the power of the highest peak found in different PSD obtained by correcting the photon arrival times for a first  period derivative component with values in the $-11<\mathrm{Log}\dot{P}<-5$ range. Colors mark the Leahy power estimates (for 2 degree of freedom) in the corresponding PSD (see intensity scale on the right). Right panel: SOPA plot of \src\ for the same data set, which was obtained by correcting the photon arrival times for a Doppler effect originated by an orbital motion. We show the orbital period as a function of $a_\mathrm{X} \sin i$.}
\label{fig:PASTA}
\end{figure*}

\subsection{Pulsation Discovery}
\label{sec:discovery}

For all time series with at least 5,000 counts, we performed an accelerated search for signals with our PASTA (Pulsation Accelerated Search for Timing Analysis, to be released in a future date) code. PASTA corrects the ToA of each photon accounting for shifts corresponding to period derivatives in the range \mbox{$-10^{-6}<\dot{P}/P[\mathrm{s}^{-1}]< 10^{-6}$}, and then looks for peaks above a self-defined detection threshold in the corresponding power spectral density (PSD) even in presence of non-Poissonian noise components \citep{israel96}. 
\begin{deluxetable*}{lcrccccc}
\tablecaption{The \xmm\ observations of \src. Uncertainties in the measurements are reported at 1$\sigma$ confidence level. \label{tab:detections}}
{\small
\tablehead{
\colhead{Data set}  & \colhead{Start date} & \colhead{pn Events\tablenotemark{a}} & \colhead{Period} &
\colhead{PF
} & \colhead{$T_{\rm Obs}$\tablenotemark{b}} & \colhead{Off-axis angle} &\colhead{EPIC ($\Delta t$)\tablenotemark{c}} \\
\colhead{Obs.ID} & \colhead{(MJD)} & \colhead{(\#)} & \colhead{(s)} &
\colhead{(\%)} & \colhead{(ks)} & \colhead{(arc min)}& \colhead{(s)}
}
\decimals
\startdata
0112840201 & 52654 & 1241&--  & $<$37 & 20.9 (19.0) & 2.3 & pn\,(0.073)\\
0212480801 & 53552 & 6140&3.2831(2)\tablenotemark{d}  & 12(2) & 49.2 (47.3) & 3.4 & pn\,(0.073)\\
0303420101 & 53875 & 5771&--  & $<$17 & 54.1 (44.0) & 3.5 & pn\,(0.073)\\
0303420201 & 53879 & 6078&--  & $<$16 & 36.8 (34.9) & 3.5 & pn\,(0.073)\\
0677980701 & 55719 & 1206&--  & $<$37 & 13.3 (11.4) & 3.6 & pn\,(0.073)\\
0677980801 & 55723 & 211&2.8014(7)\tablenotemark{d}   & 82(17)  & 13.3 (11.4) & 3.6 & pn\,(0.073)\\
\hline
0824450901 (A)& 58251 & 15082& 2.79812(5)& 15(2) & 78.0 (74.8) & 0.0 & pn\,(0.073)\\
0830191401 (L)& 58263 & 1037&--         & $<$28\tablenotemark{e}   & 98.0 (94.7) & 0.0 & pn\,(0.073),\,MOS\,(0.3)\\
0830191501 (B)& 58282 & 12083 & \multirow{2}{*}{2.7977148(2)\tablenotemark{f}}&8(2) & 63.0 (59.8) & 0.0 & pn\,(0.073),\,MOS\,(0.3)\\
0830191601 (C)& 58284 & 12559 & &10(2)  & 63.0 (59.8) & 0.0 & pn\,(0.073),\,MOS\,(0.3)\\
\enddata
\tablenotetext{\tablenotemark{a}}{Event numbers refer to the total number of photons within the source extraction region for the EPIC pn.}
\tablenotetext{\tablenotemark{b}}{Duration of the observation. In parentheses the effective EPIC-pn detector exposure time.}
\tablenotetext{\tablenotemark{c}}{The number in parentheses indicates the time resolution of the camera(s) used for the timing analysis.}
\tablenotetext{\tablenotemark{d}}{Note that the maximum period variation induced by the orbital motion (not corrected for here) is $\Delta P\sim1$\,ms. }
\tablenotetext{\tablenotemark{e}}{Events from both EPIC pn and MOSs have been used in order to infer this upper limit.}
\tablenotetext{\tablenotemark{f}}{For observations 0830191501 and 0830191601 the timing parameters were  calculated together, see the text for details.}}
\end{deluxetable*}
The threshold of 5,000 counts is a conservative value based on the relation which links the number of counts to the minimum detectable signal PF in fast Fourier transforms (FTTs), which is given by $\mathrm{PF}=\left\{\left[\frac{P_{j}}{2M}\right] \frac{4}{0.773 N_{\gamma}} \frac{(\pi j/N)^{2}}{\sin^{2}(\pi j/N)}\right\}^{1/2}$, where $P_j$ is the power in the $j$-th Fourier frequency, $N_\gamma$ and $N$ the number of counts and bins in the time series, and $M$ the number of averaged FFTs in the final PSD  (in our case $M=1$; the formula has been derived from \citealt{leahy83}). 

\begin{figure*}
\centering
\resizebox{.99\hsize}{!}{\includegraphics{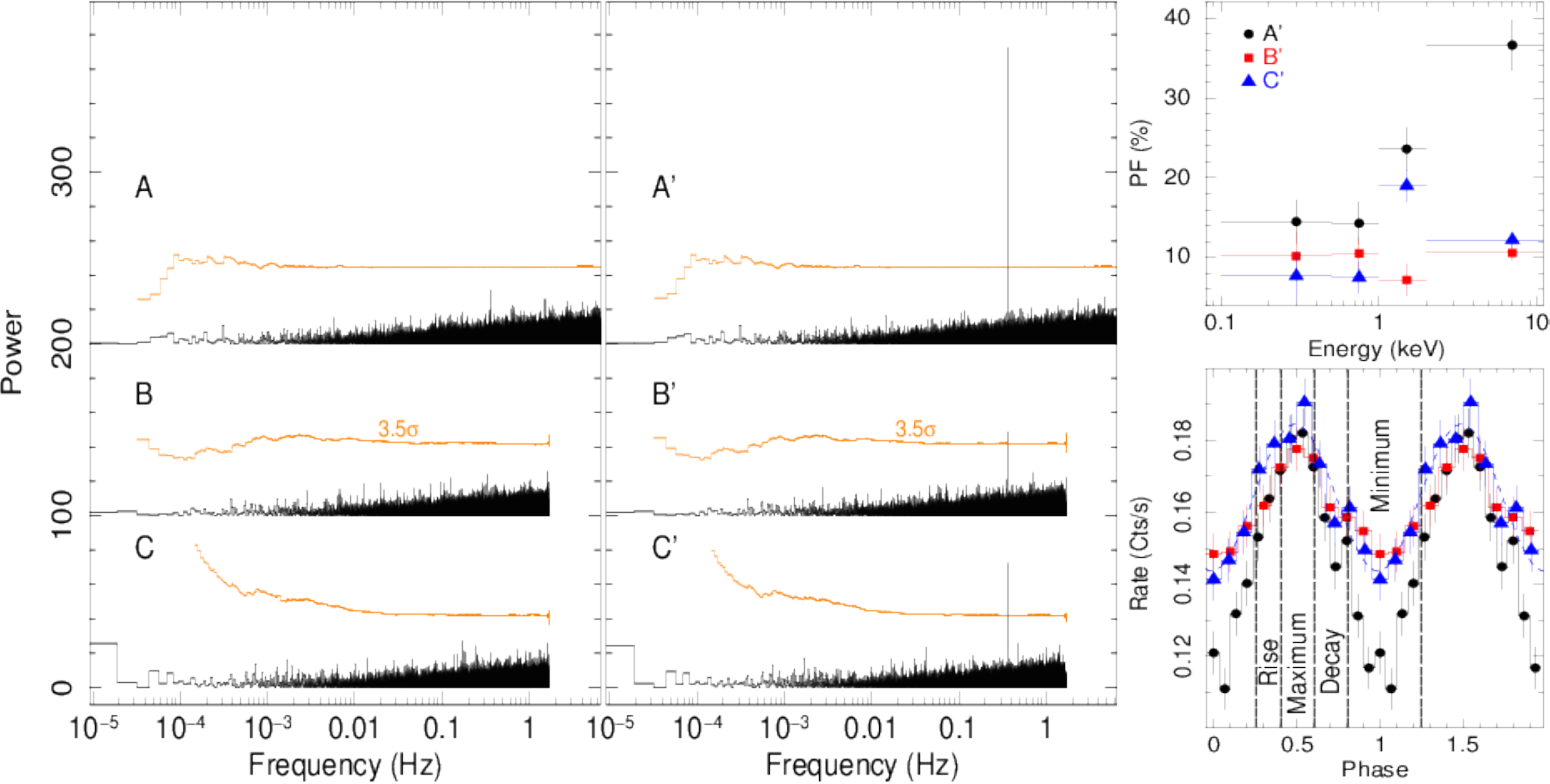}}
\caption{The power spectra of the  0.1--12.0\,keV source original \xmm\ light curves, arbitrarily shifted along the y axis, are shown in the leftmost plot together with the 3.5$\sigma$ detection threshold: 2018-05-13 (A; pn), 2018-06-13 (B; pn + MOSs), and 2018-06-15 (C; pn + MOSs). In the central panel we show the power spectra of the same light curves after correcting their photon arrival-times for both a first period derivative and an orbital Doppler term (A',B' and C'). The PFs and the pulse shapes are shown in the rightmost panels for the corrected light curves A',B' and C', as a function of the energy band. The vertical lines in the right-bottom panel indicate the phase intervals used in the pulse-resolved spectral analysis, see Sect.\,\ref{section:spectra}.}
\label{fig:timing}
\end{figure*}

Among the fifteen M51 sources for which we searched for signals with PASTA we found that only \src\ shows a relatively strong signal at $\sim$2.8 s with PF of $\sim$10\% (dataset A in Table\,\ref{tab:detections} and Fig.\,\ref{fig:PASTA} left panel; typical PASTA PF upper limits for the other brightest ULXs in M51 during the same observation are in the 7--20\% range). 
PASTA hinted at a $\dot{P} \sim 9.7 \times 10^{-8}\text{ s s}^{-1}$ for which the signal showed a power of about 80 in the corresponding PSD (see left and central upper  panels of Fig.\,\ref{fig:timing}). Based on the fact that all other PULXs are in a binary system, we assume the same is valid for \src. This implies that the observed $\dot{P}$ may not be solely due to the NS intrinsic spin $\dot P_{\mathrm int}$, but instead to the superposition of the  $\dot{P}_{\mathrm{int}}$ and an apparent, local $\dot{P}_{\mathrm{orb}}$ caused by the motion of the NS around the center-mass of the binary system (assuming an orbital inclination $ i > 0\degree$). We then used our SOPA (Search for Orbital Periods with Acceleration) code that performs a similar correction of the ToA as in PASTA but, instead of testing a grid of values for $\dot{P}$, it corrects for a set of values of the orbital parameters, assuming a circular orbit. In a first run with a sparse parameters grid, we obtained a first-order value for the orbital period $P_{\mathrm{orb}} \sim 2$\,d and a projection of the semi-major axis $a_{\mathrm{X}} \sin i \sim 25$\,lt-s.


\subsection{Timing Analysis}
We scanned the parameter space describing a pulsar in a circular orbit
around the candidate 2.8\,s period signal in the datasets of each single \xmm\
observation. We use only pn data unless the observing
mode of the MOS has a frame time shorter than the spin period, in which case we
combine all EPIC data. We found a strong signal only in the UNSEeN observation and 2 subsequent DPS observations, taken in May 13, June 13, and June 15, 2018, correspondingly (observation IDs 0824450901,
0830191501, 0830191601, labelled in the following as A, B and C, respectively). Using a direct likelihood technique as described
in \citealt{israel17} (based on \citealt{bai92} and \citealt{cowan11}), we
generated confidence profiles for the orbital parameters in each single
observation. Then we combined these results into a single ephemeris, which locks the orbital parameters between observations and allows
for two distinct sets of spin parameters ($P$ and $\dot P$): one for the first observation
in May (A), the other for the pair of observations in June (B, C). All the uncertainties
reported in this section have a confidence level of 1$\sigma$.

\begin{figure*}[th]
\centering
\resizebox{.48\hsize}{!}{\includegraphics{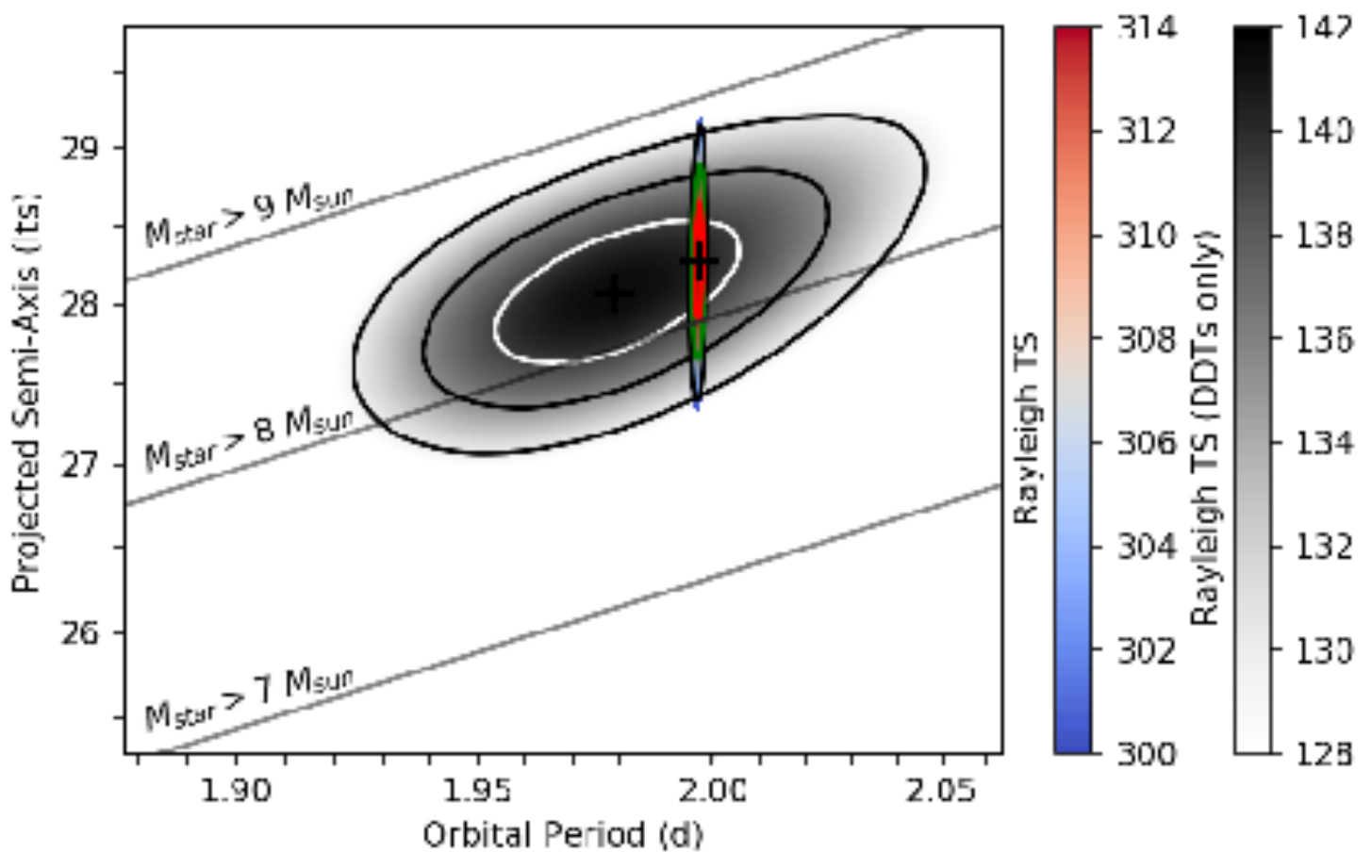}}
\hspace{3mm}
\resizebox{.48\hsize}{!}{\includegraphics{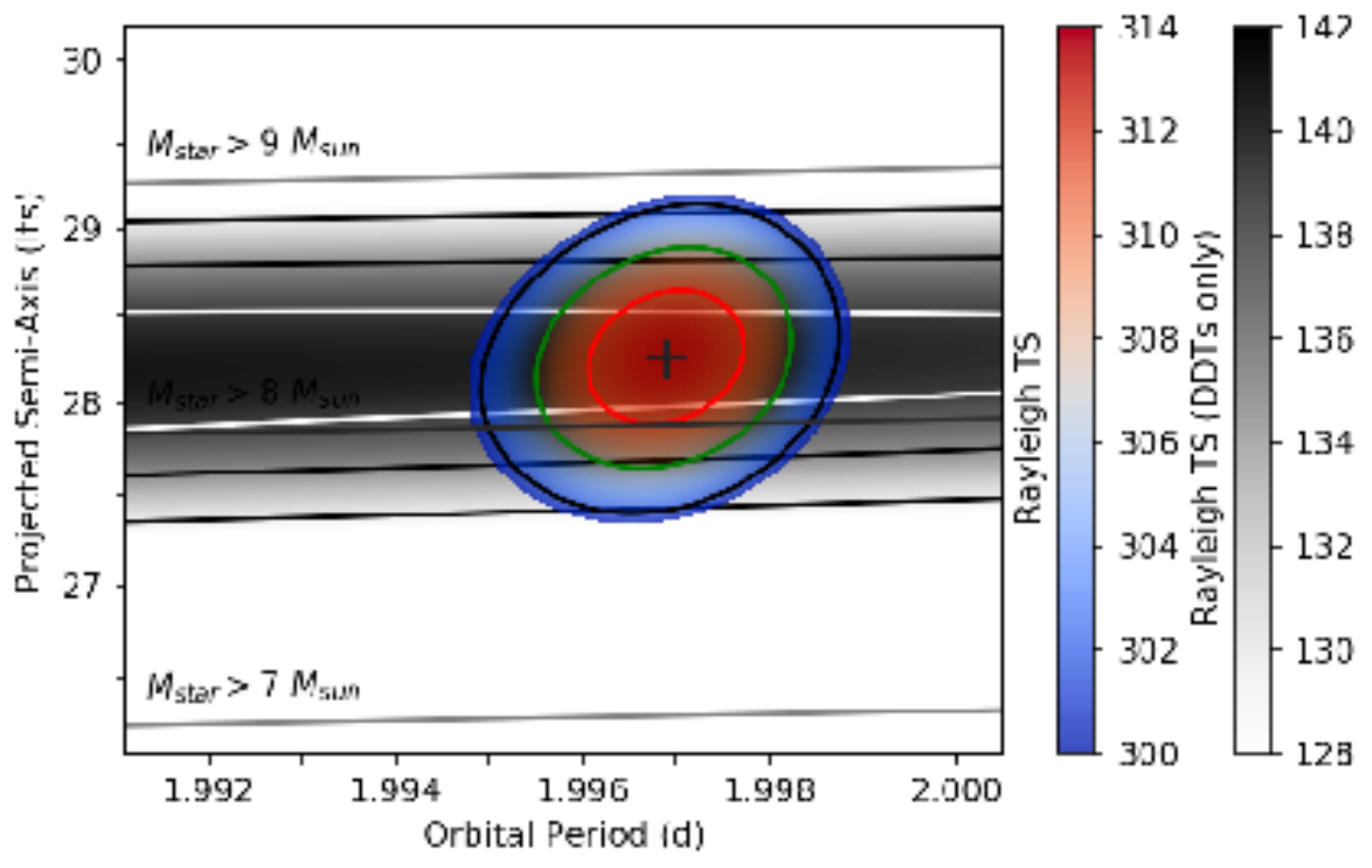}}
\caption{
Contour levels of the Rayleigh test statistics in the $a\sin i$-$P_{\mathrm orb}$ plane. The colour (gray) scale refer to  observations A+B+C (B+C) ; solid contours mark the 1, 2 and 3 $\sigma$ confidence levels. The best estimates of the projected semi-major axis and of the period are marked with a cross. Right panel. A blow-up of the same plot around the best fit values.
The solid parallel lines indicate particular configurations for 
which the orbital inclination and the masses of the two objects are 
held fixed and assuming a system observed edge-on. As the inclination of 
the system is unknown, these values 
represent lower bounds to the actual mass of the companion star.}
\label{fig:orbit}
\end{figure*}

Figure\,\ref{fig:orbit} shows the orbital period ($P_{\mathrm{orb}}=1.9969(7)\text{ d}$)
and the projected semi-major axis of the NS orbit, $a_{\mathrm{X}}\sin i=28.3(4)$\,\mbox{lt-s}, 
resulting from this coherent direct likelihood analysis. To complete
our description of the circular orbit of the system we estimated the
epoch of ascending nodes as $T_{\mathrm{asc}}=58267.036(6)\text{ MJD}$. 
Unfortunately, the signal is not strong enough to make a further step and probe an elliptical
orbit; using the binary model ELL1 \citep{lange01} in  TEMPO2 
\citep{hobbs06}, we set an upper limit on the eccentricity of the orbit of
$e<0.22$ at a 2$\sigma$ confidence level. We stress that while the orbital period inferred for \src\ is close to that of the revolution of \xmm\ around the Earth (1.994 days), the projected semi-major axis ($a_{\mathrm{X}} \sin i \sim0.5$\,\mbox{lt-s}) as well as the eccentricity ($e\sim0.7$) of the spacecraft orbit are so different from those of \src\ that we can confidently exclude any relation to it.
As we do not see eclipses, the inclination of the system is essentially unconstrained. 
Figure\,\ref{fig:orbit} shows the lower limits on the mass of the companion star for an orbit seen edge-on. Assuming a $1.4\,M_{\odot}$ neutron star and the orbital parameters in Table\,\ref{table:orbit}, $\mathbf{M_\star>8.3\,M_{\odot}}$. An upper bound on the mass of the companion star can be placed by considering that for a random distribution of orbital inclination angles over the $[0,\frac{\pi}{2}]$ interval, the probability of having an angle $i\lesssim26\degree$ is only 10\%. Therefore, we obtain $M_\star\lesssim80M_{\odot}$ at the 90\% confidence level. Similarly, for the average value of the sine function, we find $M_\star\simeq\mathbf{27}M_{\odot}$.

If we fix the circular orbit to its best-fit solution, then the spin parameters,
$P$ and $\dot{P}$ with a reference epoch at the center of each data set,
are strongly constrained, with extremely small uncertainties. However,
the spin parameters of both data sets are strongly covariant with the
orbital parameters so the stated uncertainties are obtained considering
them as nuisance parameters, through profile likelihood \citep{murphy00}. 
In the first observation, 
we obtain $P^{\left(a\right)}=2.79812(5)\text{ s}$ and 
$|\dot{P}^{\left(a\right)}|<3\times10^{-9}\text{ s s}^{-1}$ with
reference epoch $T_{0}^{\left(a\right)}=58252.36734\text{ MJD}$.
In the second set of observations,
we obtain $P^{\left(b\right)}=2.79771475(19)\text{ s}$ and
$\dot{P}^{\left(b\right)}=-2.4(6)\times10^{-10}\text{ s s}^{-1}$ with
reference epoch $T_{0}^{\left(b\right)}=58283.44400\text{ MJD}$. 

\begin{table}
\caption{Timing solutions for \src. \label{table:orbit}}
\begin{tabular}{lrr}
\hline
Parameters & A & B\,+\,C \\
\hline
Epoch $T_0$ (MJD) & 58252.36733583 & 58283.4440025\\
Validity (MJD-58251) & 0.92--1.79 & 31.09--33.78\\
$P(T_0)$(s) & 2.79812(7) &2.79771475(25)\\
$|\dot{P}(T_0)|\times{\rm 10}^{-10}$ & $<$26 &$-$2.4(7)\\
$\nu(T_0)$ (Hz) & 0.357383(9) &0.35743458(3)\\
$|\dot{\nu}(T_0)|\times{\rm 10}^{-11}$ (Hz s$^{-1}$) & $<$35 & 3.1(8)\\
\hline
$P_{\mathrm{orb}}$ (d) & \multicolumn{2}{c}{1.9969(7)} \\
$a_{\mathrm{X}} \sin i$ (lt-s) & \multicolumn{2}{c}{28.3(4)}\\
$T_{\mathrm{asc}}$ (MJD) & \multicolumn{2}{c}{58285.0084(12)} \\
$e$ & \multicolumn{2}{c}{$<$0.22 } \\
Mass function ($M_\odot$) & \multicolumn{2}{c}{6.1(3)} \\
\textbf{Min. c}ompanion mass ($M_\odot$) &\multicolumn{2}{c}{$8.3(3)$} \\
\hline
\end{tabular}
\tablecomments{Figures in parentheses represent the uncertainties 
in the least significant digits and are all at a confidence level
of 1$\sigma$. The upper limit on the eccentricity ($e$) is at 2$\sigma$.
The \textbf{minimum} companion mass is computed for a NS of $1.4\,M_{\odot}$.}
\end{table}

\subsection{Signal properties}
\label{sec:signal}
In order to characterize better the 2.8\,s pulsations, we study the time-resolved PFs of the three 2018 observations as a function of the orbital phase.  


\begin{figure}[th]
\hspace{-5mm}\includegraphics[width=7.3cm,angle=270]{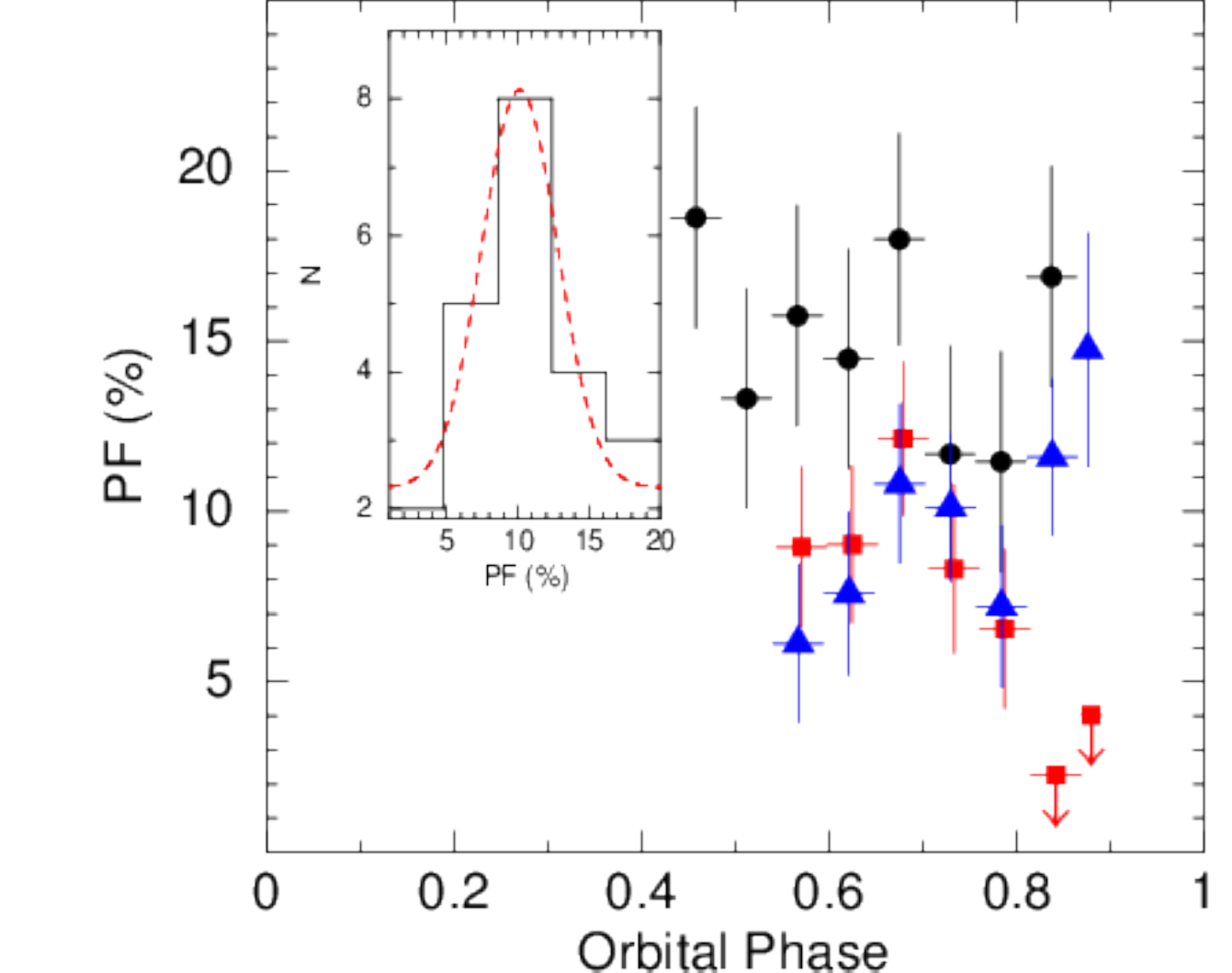}
\caption{The evolution of the PF of \src\ is shown as a function of the orbital phase for observations A, B and C (black filled circles, red squares and blue triangles, respectively). The histogram in the inset shows the distribution of the observed PFs, where N is the number of times the PF falls within a given percentage range (bin).}
\label{fig:signal}
\end{figure}

Figure\,\ref{fig:signal} shows that the PF within each observation is variable and covers a range between about $\sim$18\% and 5\%. The trend is almost constant during observation A, decreasing in B and increasing in C. Moreover, during the latest part of observation B the signal is not detected with a 3$\sigma$ upper limit of about 3\%. This variability seems to be uncorrelated with the orbital phase (signal upper limits in data set B are located at the same orbital phase where pulsations with $\rm PF>10$\% are inferred in A and C) suggesting that its origin should not be ascribed to a recurrent (orbital) geometrical effect. The PF distribution for the three data sets suggests that values below 5\% are possible, but remain undetected due to low statistics (see the inset in Fig.\,\ref{fig:signal}). 

While the pulse shape remains unchanged among pointings, within uncertainties, the PFs dependence on the energy is only partially similar to that observed in other PULXs (see Fig.\,\ref{fig:timing} right panels). In fact, the typical increase of the PF for increasing energies is only observed during observation A, while the PFs remain almost constant in observation B and C, making \src\ the first PULX showing a time-dependent behaviour of the PF as a function of energy.   
We also note that the 0.1--12 keV PFs seem to be inversely proportional to the source flux (see Table\,\ref{tab:detections} and right plots of Fig.\,\ref{fig:timing}) which  is in agreement with the idea that the soft thermal component ({\sc diskbb}), the only component which is changing among observations A, B and C, 
is less pulsed (see Sect.\,\ref{section:spectra}). 
In all the three observations (A, B and C) the pulse profile at high energies ($E>1$\,keV) precedes that at lower energies by  0.1--0.2 in phase, similar to the case of NGC5907\,ULX1 \citep{israel17}. 

Concerning $\dot{P}$, we note that while the second set of observations (B, C) indicates a spin-up trend, during the first observation (A) the intrinsic $\dot{P}$ could not be constrained. Comparing the spin period at the two epochs, roughly one month apart, we obtain a spin-up rate $\dot{P}=-1.5(1)\times10^{-10}\text{ s s}^{-1}$. This is marginally compatible  with both the larger spin-up rate measured during 
the second epoch, $\dot{P}^{\left(b\right)}$, and with the $\dot{P}^{\left(a\right)}$ upper limit  inferred during  the first epoch.
This might be a mild indication
that the source might have partially entered a propeller spin down phase while it was in a low state during observation L (Obs.ID 0830191401) between pointings A and B (see also Sec.\,\ref{sec:discussion}).

\begin{figure}[tb]
\centering 
\hspace{-0mm}\includegraphics[width=8.2cm,height=9cm,angle=270]{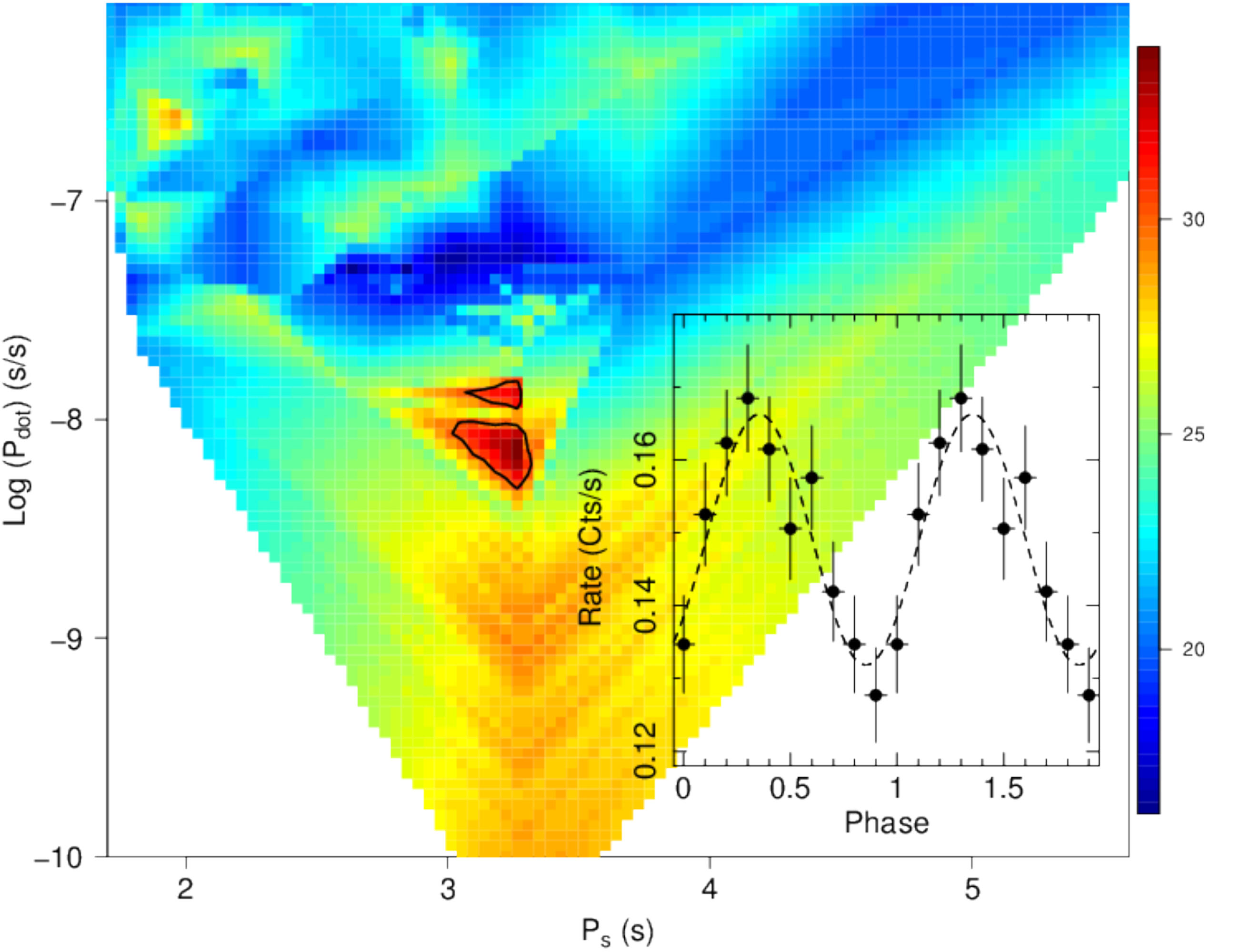}
\caption{PASTA plot for the archival data set 0212480801 of \src\ (see caption of Fig.\,\ref{fig:PASTA} for more details). It shows the map of the estimated power, peaking at about $P=3.28$\,s and  $\dot{P}\sim10^{-8}$\,s\,s$^{-1}$. The light curve folded at the best inferred period is superimposed.}
\label{fig:old}
\end{figure}
\begin{figure*}
\center
\includegraphics[width=6.0cm,angle=270]{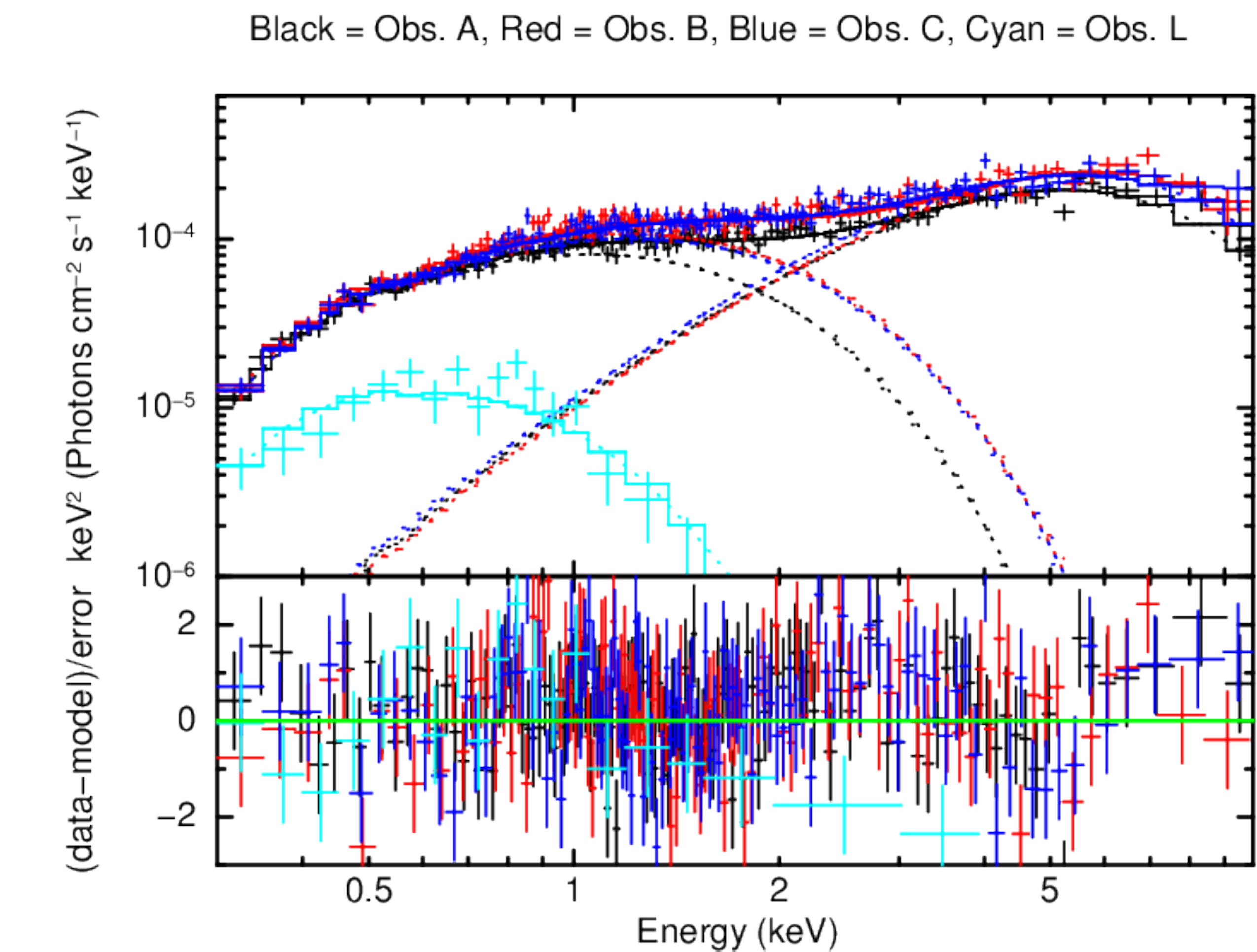}
\includegraphics[width=6.0cm,angle=270]{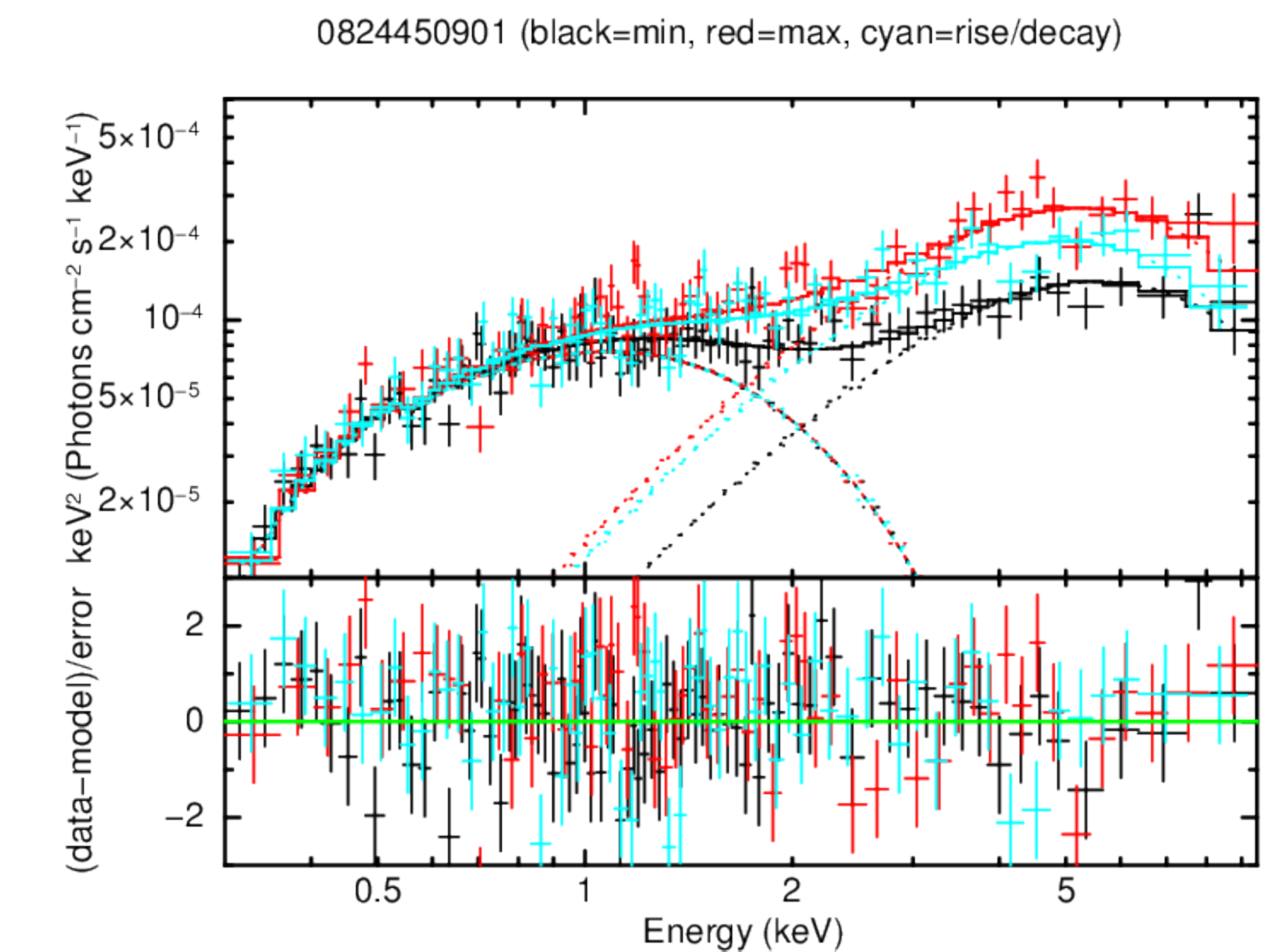}
\includegraphics[width=6.0cm,angle=270]{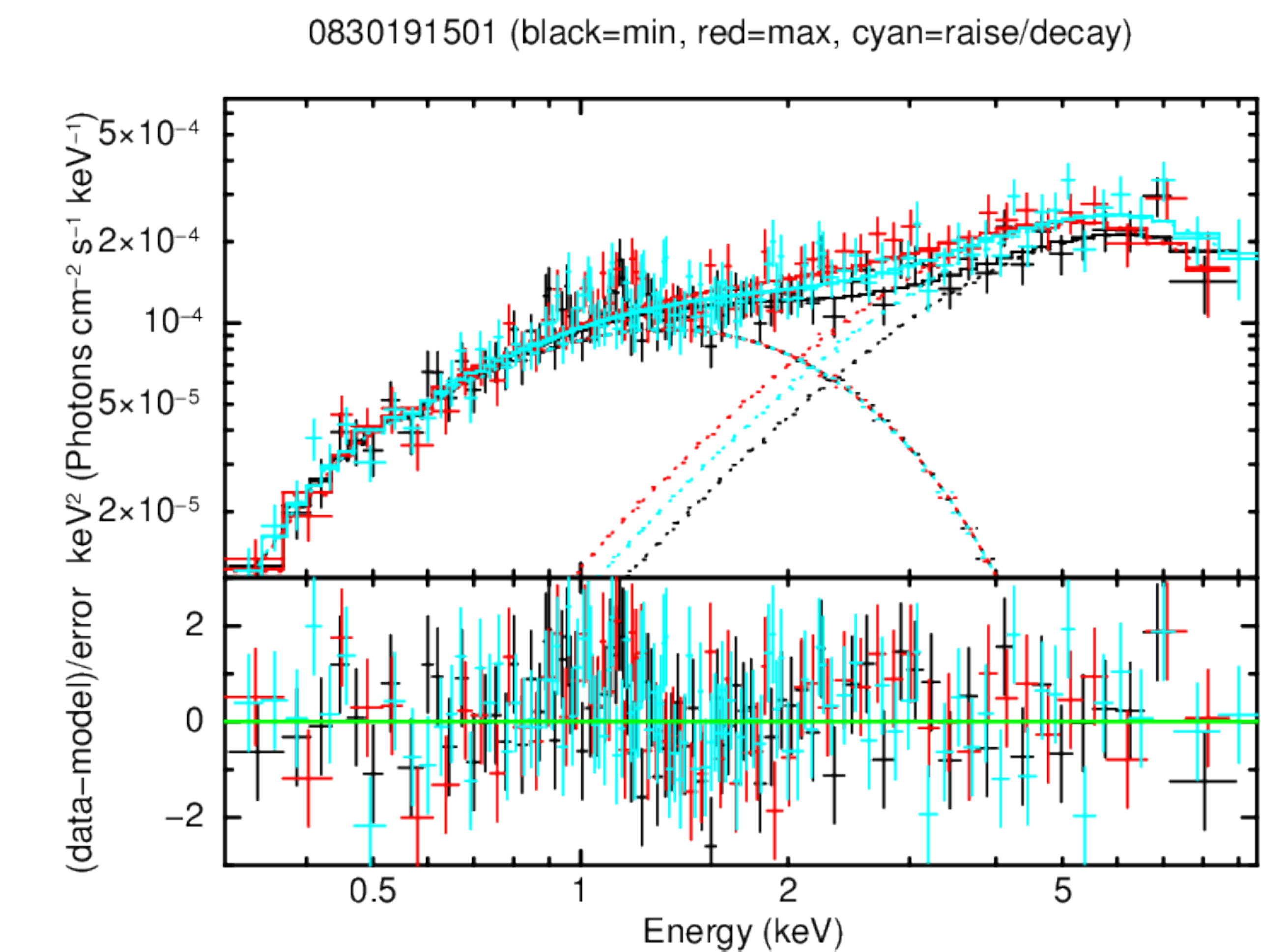}
\includegraphics[width=6.0cm,angle=270]{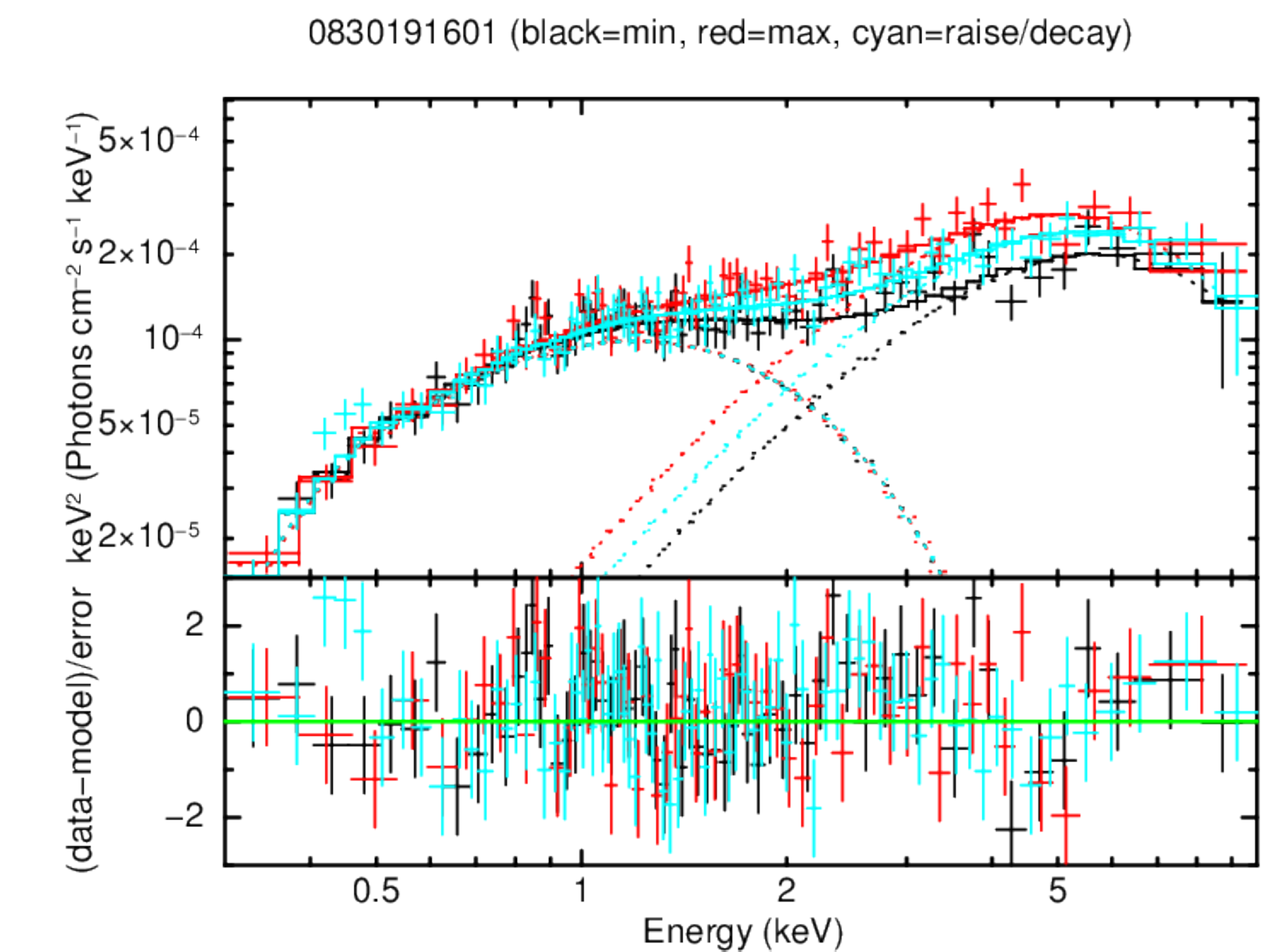}
\caption{Unfolded spectra ($E^2f(E)$) of the four \xmm\ observations (where the color code is black: observation A, cyan: observation L, red: observation B, blue: observation C), fitted with a {\sc tbabs(diskbb+bbodyrad)} model (upper left panel). Phase-resolved spectra of observation A (upper right), B (bottom left) and C (bottom right) fitted with the same model used for the phase averaged spectra. 
See Section\,\ref{section:spectra} for more details about the phase-resolved analysis. Spectra and residuals have been further rebinned for visual purposes only.}
\label{phase_spec}
\end{figure*}

\subsection{Archival data}

We analysed all the six \xmm\ archival data sets of \src\ in order to check for the presence of a signal consistent with 2.8\,s (see Table\,\ref{tab:detections}). In particular, we run the three timing techniques described above to find and to study any significant signal. We detected the spin modulation in two observations (in one of the two with marginal significance) taken in 2005 (021280801) and 2011 (0677980801), both at 5$\sigma$ (single trial) and with formal $\dot{P}\sim10^{-8}\text{ s s}^{-1}$ and $\sim$\,$10^{-7}\text{ s s}^{-1}$, respectively.  
Taking into account the number of trials, which was estimated as the ratio
between the offset in spin frequency from our detections in 2018 and the 
intrinsic Fourier resolution of each exposure, the significance of the 
signal goes down to 4.7$\sigma$ and 3.3$\sigma$ in the 2005 and 2011 
observations, respectively. An inspection of the timing properties  of the candidate signals strongly suggests that the signal detected in the 2011 (55723 MJD) data set is likely spurious (highly non-sinusoidal pulse shape with an abnormally high PF of 82\% at variance with the signal properties reported in the above sections; see Table\,\ref{tab:detections}). 
In addition, the 2011 source flux is similar to that observed in deeper observation L, where we did not find any signal down to an upper limit of $\sim$30\,\% (see Table.\,\ref{tab:detections}). Therefore, we considered this detection as not reliable. For the 2005 detection, we cannot disentangle the orbital contribution from the measured spin period $P=3.2831$\,s and its time derivative $\dot{P}$, though we can estimate that the maximum $P$ variation induced by the orbital parameters is only of the order of 1\,ms, while the 2005 period is $\sim 0.4$ s longer than that in 2018.
Correspondingly, we obtain an average first period derivative of $\dot{P}\simeq -10^{-9}\text{ s s}^{-1}$, a value in the range of the $\dot{P}$ observed in the other PULXs. As this was obtained from a long baseline (13 years), it is virtually unaffected by the orbital Doppler shift (which is instead present in each single observation) and, therefore, can be considered as a good estimate of the secular accretion-induced $\dot{P}_{\mathrm{sec}}$.


\subsection{Spectral analysis}
\label{section:spectra}



We first simultaneously fitted the average
spectra of each of the four 2018 observations with a model commonly adopted for PULXs spectra in the 0.3-10 keV range, consisting of
a soft component (a multicolour blackbody disc; {\sc diskbb};
\citealt{mitsuda84}) plus a higher-temperature blackbody ({\sc bbodyrad}), both absorbed by a total covering column ({\sc tbabs})\footnote{We  verified that, in observation B, the source did not spectrally vary during the `on' and `off' phases and therefore we analyzed the whole average source spectrum.}. Abundances were set to that of \cite{wilms00} and cross section of \cite{verner96}. 
We note that the \xmm+{\it NuSTAR} spectra of a sample of bright ULXs (comprising two PULXs, \citealt{walton18b}) were fitted with the same model plus a {\sc cutoffpl} component, where the latter was associated with the emission of an accretion column. We have verified that the inclusion of this extra component in our data marginally improved the fit, although its spectral parameters were poorly constrained, because of the smaller energy range. Hence, we proceeded to use only the two-components model. 
Initially, we let all spectral parameters free to vary and we obtained a statistically acceptable fit 
($\chi^2_{\nu}/$dof$ =1.05/1\,739$). 
We noticed that the $N_{\rm H}$ was consistent within uncertainties with a constant value across different epochs and that the fit was insensitive to changes of the parameters of the {\sc bbodyrad} component for the lowest flux observation (L). Therefore, its normalization was set to zero and a new fit was performed by linking $N_{\rm H}$ across all observations (e.g. assuming that the column density
did not change significantly between the different epochs). The average spectra best fit has $\chi^2_{\nu}/$dof$ =1.06/1\,744$, the best fit model parameters are reported in Table~\ref{average_spec} and spectra are shown in Fig.~\ref{phase_spec} (upper left panel).
\begin{table*}
\caption{Best fit to the average 0.3--10\,keV spectrum using {\sc tbabs(diskbb + bbodyrad)}. 
The fluxes and the luminosity are also reported (in the 0.3--10\,keV band). 
Uncertainties are at $1\sigma$ confidence level.}
{\small
\begin{center}
\begin{tabular}{ccccccccccc}
\hline 
               &\multicolumn{1}{c}{{\sc tbabs}} & \multicolumn{3}{c}{{\sc diskbbody}} & \multicolumn{3}{c}{{\sc bbodyrad}}  & \multicolumn{1}{c}{} & \multicolumn{1}{c}{} & \multicolumn{1}{c}{}  \\
Obs.            & \multicolumn{1}{c}{$N_{\mathrm{H}}$}     & \multicolumn{1}{c}{$kT_{\rm in}$}        & \multicolumn{1}{c}{Radius$^{a,b}$} & \multicolumn{1}{c}{Flux$_{\rm dbb}^*$}         & \multicolumn{1}{c}{$kT$}            &  \multicolumn{1}{c}{Radius$^b$}   & \multicolumn{1}{c}{Flux$_{\rm bb}^*$}                      & \multicolumn{1}{c}{$F_{\rm X}^+$}              & \multicolumn{1}{c}{$L_{\rm X}$ $^b$} &   \multicolumn{1}{c}{$\chi^2$/dof}  \\
               & \multicolumn{1}{c}{($10^{20}$ cm$^{-2}$)}           &           (keV)     &   \multicolumn{1}{c}{ (km)}      &     \multicolumn{1}{c}{}       &      (keV)         & \multicolumn{1}{c}{ (km)}   &             & \multicolumn{1}{c}{}              & \multicolumn{1}{c}{($10^{39}$\,\lum)}      &                \\
\hline 
A & \multirow{4}{*}{$5.9^{+0.6}_{-0.5}$} & $0.40^{+0.01}_{-0.01}$ & $867^{+58}_{-58}$ & $2.34^{+0.05}_{-0.05}$ & $1.33^{+0.04}_{-0.03}$ & $96^{+4}_{-4}$ & $4.01^{+0.08}_{-0.08}$ & $5.9^{+0.1}_{-0.1}$ & $5.6_{-0.1}^{+0.1}$       &    \multirow{4}{*}{$1.06/1\,744$}  \\
L & & $0.19^{+0.01}_{-0.01}$ & $1673^{+26}_{-22}$ & $0.33^{+0.02}_{-0.02}$ & (1.33 fix.) & --  & $<0.03^c$ & $0.21^{+0.01}_{-0.01}$ &  $0.29_{-0.02}^{+0.02}$ &  \\
B & & $0.47^{+0.02}_{-0.02}$ & $728^{+49}_{-42}$ & $3.10^{+0.07}_{-0.07}$ & $1.50^{+0.05}_{-0.05}$ & $86^{+5}_{-5}$ & $4.94^{+0.12}_{-0.12}$ & $7.5^{+0.1}_{-0.1}$ & $7.1_{-0.1}^{+0.1}$  & \\
C & & $0.47^{+0.02}_{-0.02}$ & $697^{+51}_{-51}$ & $2.93^{+0.07}_{-0.07}$ & $1.40^{+0.05}_{-0.04}$ & $97^{+6}_{-6}$ & $4.85^{+0.11}_{-0.11}$ & $7.3^{+0.1}_{-0.1}$ & $6.8_{-0.1}^{+0.1}$  & \\
\hline
\end{tabular}  
\label{average_spec}                      
\end{center}} 
\begin{flushleft}
$^*$ 0.3--10 keV unabsorbed fluxes in units of $10^{-13}$\,\flux. \\
$^+$ 0.3--10 keV absorbed fluxes in units of $10^{-13}$\,\flux. \\
$^a$ Assuming an inclination angle of 60\textdegree\ and not correcting for a color factor. \\
$^b$ Assuming $d=8.58$\,Mpc.\\
$^c$ 3$\sigma$ upper limit. \\
\end{flushleft}                               
\end{table*}

\begin{table*}
\caption{Best fit to the phase-resolved 0.3--10 keV spectra obtained using a {\sc tbabs(diskbb+bbodyrad)} model.
Uncertainties are at $1\sigma$ confidence level.}
{\small
\begin{center}
\begin{tabular}{lccccccc}
\hline 
& &\multicolumn{1}{c}{{\sc tbabs}} & \multicolumn{2}{c}{{\sc diskbbody}} & \multicolumn{2}{c}{{\sc bbodyrad}}  & \multicolumn{1}{c}{}    \\
               \hline
Obs.    & Phase        & \multicolumn{1}{c}{$N_{\rm H}$}     & \multicolumn{1}{c}{$kT_{\rm in}$}        & \multicolumn{1}{c}{Norm}         & \multicolumn{1}{c}{$kT$}            &  \multicolumn{1}{c}{Radius$^a$}                        &   \multicolumn{1}{c}{$\chi^2_{\nu}$/dof}  \\

    &           & \multicolumn{1}{c}{($10^{20}$\,cm$^{-2}$)}           & \multicolumn{1}{c}{(keV)}           &                      & \multicolumn{1}{c}{(keV)}           &        \multicolumn{1}{c}{(km)}                     &                \\        
\hline 
\multirow{3}{*}{A} & min. & \multirow{3}{*}{$5.9^{*}$} & \multirow{3}{*}{$0.41^{+0.02}_{-0.02}$}  & \multirow{3}{*}{$0.45^{+0.06}_{-0.05}$}  &  $1.43^{+0.09}_{-0.08}$  & $71_{-6}^{+8}$ & \multirow{3}{*}{$1.07/467$}     \\
  & raise/decay & & & & $1.27^{+0.06}_{-0.05}$  & $107_{-8}^{+7}$ &     \\
  & max.  & & & & $1.35^{+0.05}_{-0.05}$  & $110_{-7}^{+7}$ &     \\
  \hline
\multirow{3}{*}{B} & min. & \multirow{3}{*}{$5.9^{*}$} & \multirow{3}{*}{$0.53^{+0.02}_{-0.02}$}  & \multirow{3}{*}{$0.18^{+0.02}_{-0.02}$}  & $1.57^{+0.09}_{-0.08}$  & $72_{-8}^{+8}$ & \multirow{3}{*}{$1.10/437$}     \\
  & raise/decay   & & & & $1.52^{+0.07}_{-0.07}$  & $84_{-6}^{+6}$ &     \\
  & max. & & & & $1.34^{+0.07}_{-0.06}$  & $104_{-7}^{+7}$ &     \\
  \hline
\multirow{3}{*}{C} & min. & \multirow{3}{*}{$5.9^{*}$} & \multirow{3}{*}{$0.47^{+0.02}_{-0.01}$}  & \multirow{3}{*}{$0.33^{+0.04}_{-0.04}$}  & $1.46^{+0.08}_{-0.07}$  & $81_{-6}^{+6}$ & \multirow{3}{*}{$1.04/690$}     \\
  & raise/decay   & & & & $1.38^{+0.05}_{-0.05}$ & $99_{-7}^{+7}$ &     \\
  & max. & & & & $1.29^{+0.05}_{-0.05}$  & $122_{-8}^{+8}$ &     \\
  \hline
\end{tabular}  
\label{phase_spec_par}                      
\end{center}} 
\begin{flushleft}
$^{*}$ Fixed to the average spectra best fit value. \\
$^a$ Assuming $d=8.58$\,Mpc.\\
\end{flushleft}                               
\end{table*}

We subsequently performed a phase-resolved spectral analysis. 
We selected three phase intervals with respect to the NS spin rotation, where the hardness ratios between the 
light-curve in the energy bands \mbox{0.3--2} and 3--10\,keV showed the largest variations (in this case we excluded the time interval of observation B where pulsations are not detected). These phase intervals correspond to the minimum (0.8--1.25), maximum \mbox{($\sim$0.4--0.6)}, and 
rise/decay of the spin pulse profile ($\sim$0.6--0.8 and 0.2--0.4; see Fig.\,\ref{fig:timing}, right panels). 
For each observation, we fitted the spectra of the three phase-bins simultaneously, using the same model adopted for the average 
spectra. We fixed the column density to the average spectra best-fit value ($5.9\times10^{20}$\,cm$^{-2}$) and we left all the other parameters 
free to vary. We found that the temperature and the 
normalization of the {\sc diskbb} are constant within each observation with respect to phase changes. The temperature and normalization of the high-energy component are instead variable. For this reason, for each observation, we also linked the {\sc diskbb} temperature and normalization amongst the spectra of the three spin-phase intervals and performed a further fit. The best-fit values are reported in Table\,\ref{phase_spec_par} and the corresponding spectra are in Fig.\,\ref{phase_spec} (upper right and bottom panels). The phase-resolved spectra shows also the presence of some broad absorption residuals around 4-5 keV. We tentatively included a {\sc gabs} model to account for an absorption feature, but the improvement in the best-fit was not significant. More investigations and possibly data would be necessary to confirm or disprove their existence.

\section{Archival \emph{HST} data}\label{counterparts}

We searched for a possible optical counterpart to \src\ using archival
{\em Hubble Space Telescope (HST)} Advanced Camera for Surveys (ACS) data
from the Legacy Project targeted at M51 and its companion galaxy NGC\,5195 
(proposal ID 10452, PI S.\,Beckwith) and the most accurate source position \citep{kuntz16}.


\begin{figure}[tb]
\includegraphics[width=8.8cm]{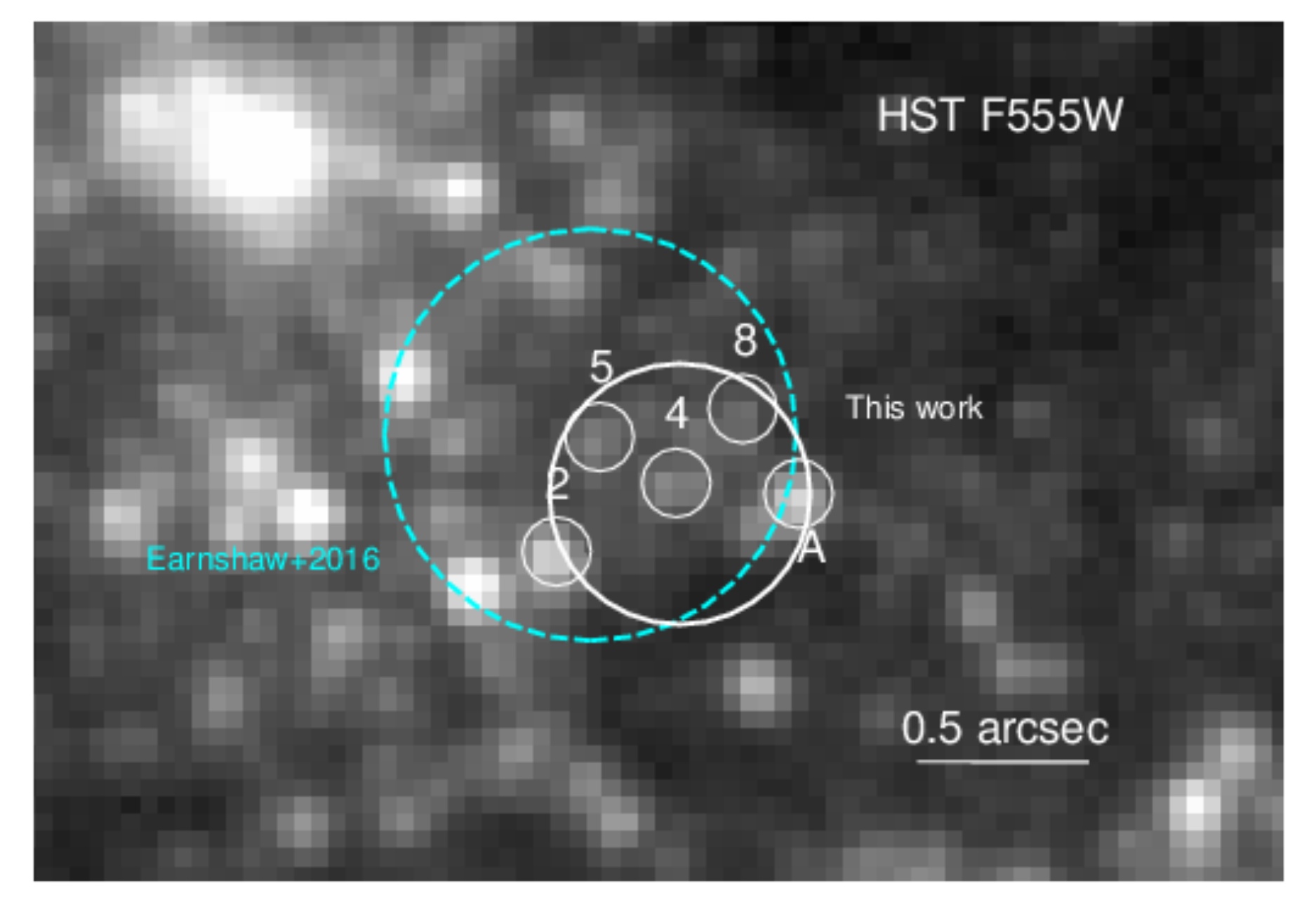}
\caption{The field of \src\ as seen by {\em HST} in the F555W band. The 
error circle is dominated by the uncertainty on the X-ray to optical image
registration and has a radius of $0\farcs38$, corresponding to 3 times 
the r.m.s. of the \chandra-to-{\em HST} source superimposition.
Source A has an absolute magnitude 
$M_V\sim-7.4$, with $B-V\sim0$; it was not considered by \cite{earnshaw16}
as a potential counterpart to \src. Objects 2, 4, 5 and 8 are numbered 
according to \cite{earnshaw16}. See text for more details. }
\label{fig:hst}
\end{figure}

We retrieved calibrated, co-added, geometrically-corrected mosaics from the
Hubble Legacy Archive\footnote{See  \url{https://hla.stsci.edu/hlaview.html}.} 
in the F435W, F555W, and F814W filters, corresponding to the B, V,  and I bands.
We first improved the absolute astrometry of {\em HST} images using 445 
sources  with a match in {\em Gaia} Data Release 2 (with an r.m.s. accuracy 
of $\sim$60\,mas). Then, we adopted the X-ray source catalogue produced by
\citet{kuntz16}. To register \chandra\ astrometry to the {\em HST} 
reference frame, we searched for close, non-ambiguous matches between 
X-ray sources and {\em HST} sources in the F814W band. We selected 19 
good reference objects, including 2 bright foreground stars (a further, 
matching foreground star was excluded from the list because of its large proper motion, 
measured by {\em Gaia}). We then aligned
\chandra\ coordinates so to match that of {\em HST}, with a resulting r.m.s. 
of $\sim0\farcs12$, in agreement with results reported by \citet{kuntz16}. 
In Figure\,\ref{fig:hst}
we show the \chandra\ error circle of \src\ on the {\em HST} F555W image, 
assuming a radius of $0\farcs38$ (3 times the r.m.s. of
the \chandra-to-{\em HST} frame registration). Our error circle is broadly
consistent with the one adopted by \citet{earnshaw16}. At least 5 {\em HST}
sources lie within our error region. The bright object marked as source A, 
not considered by \citet{earnshaw16} as a possible counterpart to \src, has
magnitudes $m_{435W}=22.4\pm0.1$, $m_{555W}=22.4\pm0.1$, and
$m_{814W}=22.5\pm0.1$, as estimated using the \textsc{SExtractor} software
\citep[][but a word of caution is appropriate in using photometric 
results, in view of the very crowded nature of the field]{bertin96}. Adopting a 
distance to M51 of 8.58\,Mpc as well as an $E(B-V)=0.030$ (following
\citealt{earnshaw16}),  its absolute magnitude is
$M_V\sim-7.4$, with $B-V\sim0$. Sources 2, 4, 5, and 8 were already 
discussed by \citet[][see their table\,6 for magnitudes 
and colours]{earnshaw16}. 

\section{Discussion}
\label{sec:discussion}
In 2018, taking advantage of the high throughput of \xmm\ and good time resolution 
of the  EPIC cameras, within the UNSEeN Large Program, we detected pulsations at a period of $\sim$2.8\,s 
in the X-ray flux of the variable source \src. We found that 
the signal corresponding to the NS spin is affected by both a secular, intrinsic, 
spin--up evolution and a Doppler effect due to the revolution 
of the pulsar around the barycenter of the binary system in an $\sim$2\,d-long orbit. All these 
findings unambiguously point to \src\ as a new member of the growing class of 
PULXs. 


During our observations and earlier campaigns
(e.g. \citealt{earnshaw16}), the (isotropic) X-ray luminosity of \src\ extended over a 
range from \mbox{$L_{\mathrm{iso,min}}\leq3\times10^{38}$} to 
\mbox{$L_{\mathrm{iso,max}}\sim10^{40}$\,\lum}, with variations occurring
on timescales longer than several days and an average luminosity level 
of \mbox{$\langle L_{\rm iso} \rangle\simeq4\times10^{39}$\,\lum}.


We detected periodic pulsations in four distinct \xmm\ observations, 
three times during May--June 2018 and once in a 2005 archival data set,
when the X-ray source luminosity was 
$L_{\mathrm{X}}\simeq (6$--$8)\times10^{39}$\,\lum.
In two cases (observations B and C), the signal was weaker   
and could be detected only after photon arrival times were corrected 
for orbital parameters using our SOPA code described in Sect.\,\ref{sec:discovery}.
The pulsed signal showed variable properties that appear to be independent 
of the orbital phase. In particular, the PF of the spin modulation decreased to 5\% within few hours from 
an initial value of about 12\%, and became undetectable 
(upper limit of about 3\%) close to the end of observation B. 
This is the first time that such strong changes in pulsation amplitude 
are observed on timescales as short as hours for a PULX. 
We note that a similar pulsation drop-out has been recently found in \nustar{} data of LMC X-4 when the source was close to the Eddington luminosity \citep{brumback18}. 
During observations B and C, the 
PF did not show the characteristic increase with energy 
that was observed in A and in other PULXs, while
the B and C X-ray spectra did not show  
significant variations above a few keV with respect to that of observation A. 


The \xmm\ spectrum of \src\ could be modelled with the sum of two thermal components, a commonly adopted model for  ULX spectra (see e.g. \citealt{pintore15,koliopanos17,walton18b}). We found that the spectral properties during observations A, B and C did not change significantly (spectral parameters broadly consistent to within 3$\sigma$) with the exception of the overall normalization;
the unabsorbed 0.3--10\,keV luminosity was $\sim$(6--8)$\times$10$^{39}$\,\lum\,(for a distance of 8.58 Mpc). The temperature of the soft component, a multicolour blackbody disc, was $\sim$0.4--0.5\,keV, in agreement with the soft temperatures observed in other ULX spectra (e.g. \citealt{gladstone09,sutton13,pintore14, middleton15}). 
The soft component may represent the emission from the inner regions of the accretion disc. The large spin-up experienced by the NS implies that the magnetic field truncates the disc at the magnetospheric radius ($r_{\mathrm{m}} \propto B^{4/7} L_{\mathrm{X}}^{-2/7}$, where $B$ is the dipole magnetic field and $L_{\mathrm{X}}$ is the X-ray luminosity).
In this scenario, the inner disc radius $R_{\mathrm{in}}$ is equal to $r_{\mathrm{m}}$, implying that $L_{\mathrm{X}}\propto K^{-7/4}$, where $K \propto R_{\mathrm{in}}^2$ is the normalization of the {\sc diskbb} model.
We fitted the best-fit $K$ and $L_{\mathrm{X}}$ values with a power law and obtained an index of $-1.6 \pm 0.4$, which is fully consistent with the expected value of $-7/4$.
The inner disk radius determined from the multicolour blackbody disc 
ranges between 700 and 1700\,km for an assumed disk inclination of 60$\degr$ (higher system inclinations are unlikely owing to the absence of eclipses or dips in the X-ray light-curve), or between 2000 and 5000 applying a colour correction factor with a typical value for this parameter $f=1.7$ \citep[e.g.][]{miller03}, that increases the size of the estimated inner disc radius by a factor $f^2$.
The harder component, which we fitted with a blackbody, had a peak temperature of $\sim$1.4\,keV  and emitting radius of $\sim$90--100\,km; the luminosity associated to it  
was about 1.6--1.7 times that of the soft component. 

The phase-resolved spectroscopy shows that the pulse variability is mainly associated with 
the harder component, which also drives the energy dependence of the PF. 
The normalization and temperature of the blackbody change with the pulse phase: the equivalent emitting radius of this component is $\sim$70 and \mbox{$\sim$110--120\,km} around the minimum and maximum of the pulse modulation, respectively. 



\subsection{Accretion model}
\label{accretion:model}
Matter accreting onto the neutron star surface releases an accretion 
luminosity of  $L_{\rm acc} (R) = GM\dot M/R \simeq 0.1 \dot M\,c^2$
(where $M$ is the NS mass, $\dot M$ the mass accretion rate, $G$
the gravitational constant, $R$ the NS radius, and $c$ the light speed).
According to the standard scenario for accreting, spinning, 
magnetized NSs, the accreting matter is able to reach the surface of the compact 
object (and hence produce pulsations at the spin period), 
when the gravitational force exceeds the centrifugal force caused by drag at the 
magnetospheric boundary. The latter is estimated based on the 
magnetic dipole field component, which at large
distances from the NS dominates over higher order multipoles. 
This condition translates into the requirement that 
the magnetosphere boundary of radius $r_{\mathrm{m}}$ is smaller 
than the corotation radius $r_{\mathrm{co}} =  \left(\frac{G M P^2}{4 \pi^2}\right)^{1/3}$  
where a test particle in
a Keplerian circular orbit corotates with the central object. 
When the above condition is not satisfied, accretion is inhibited by the centrifugal barrier  
at $r_{\mathrm{m}}$ (propeller phase) and a lower accretion luminosity 
$L_{\mathrm{acc}}{(r_{\mathrm{m}}) = GM\dot M/r_{\mathrm{m}}}$ is released.
By adopting the standard expression for the magnetospheric radius: 
\begin{eqnarray*}
r_{\mathrm{m}}& =&\frac{\xi \mu^{4/7}}{\dot{M}^{2/7}(2GM)^{1/7}}\\ 
& = & 3.3\times10^{7}\,\xi_{0.5}\,B_{12}^{4/7}\,L_{39}^{-2/7}\,R_6^{10/7}\,M_{1.4}^{1/7}~\mbox{cm} 
  \end{eqnarray*}
where $\xi$ is a coefficient that depends on the specific model of disk-magnetosphere interaction (see \citealt{campana18} for a recent estimate), $\mu$ is the magnetic dipole moment 
, $B_{12}$ is the dipolar magnetic field at the magnetic poles in units of $10^{12}$\,G, $M_{1.4}$ the NS mass in units of 1.4\,$M_{\odot}$, $\xi_{0.5}$ is the $\xi$ coefficient in units of 0.5 and $R_6$ the NS radius in units of $10^6$\,cm), the minimum accretion luminosity below which the centrifugal barrier begins to operate reads
\mbox{$L_{\mathrm{accr,min}}\simeq 4\times10^{37}\xi^{7/2}B_{12}^2P^{-7/3}M_{1.4}^{-2/3}R_6^5$\,\lum} 
(here the spin period $P$ is in seconds).
The accretion luminosity drop when the NS has fully entered the propeller regime is given by 
\mbox{$\sim$$170P^{2/3}M_{1.4}^{1/3}R_6^{-1}$} \citep{corbet96,campana00,campana01,campana02,mushtukov15,tsygankov16,campana18}. 

For the 2.8\,s spin period of \src, the accretion luminosity ratio 
across the accretor/propeller transition is expected to be a factor of about 340, {\it i.e.}   
about 10 times larger than the observed luminosity swing of 
$L_{\mathrm{iso,max}}/L_{\mathrm{iso,min}}\sim 30$ in the 0.3--10\,keV range.\footnote{We note that despite careful selection of the source and background regions, see Sect.\,\ref{data_reduction}, we cannot completely exclude some contamination in our spectra from the diffuse emission of the host galaxy. It is therefore possible at $L_{\mathrm{min}}$ (observation L) the source luminosity was slightly lower than that reported in Table\,\ref{average_spec}.} 
For the luminosity levels at which pulsations are detected,
accretion must be taking place uninhibited onto the NS surface.
One possibility is that for a fraction of the luminosity range of 
$L_{\mathrm{iso,max}}/L_{\mathrm{iso,min}}\sim 30$, the source 
has partially entered the centrifugal gap, with only a fraction of the accretion flow 
reaching the NS surface and the rest being stopped at $r_{\mathrm{m}}$ 
 (see the case of 4U\,0115+63; \citealt{campana01}). In that case, the NS would 
be close to its equilibrium period, with the magnetospheric radius close to the 
corotation radius, and undergo large variations in $\dot P$ and spin-down episodes
(see for example the case of the $\dot P$ variation of M82\,X-1; \citealt{dallosso15}), such 
that a very high secular spin-up rate would not be expected. 
This is at variance with the  secular $\dot P \sim -10^{-9}$\,s\,s$^{-1}$ of \src. 
Consequently, it is likely that the source was in the accretion regime over the  
entire luminosity swing so far observed. 

\begin{figure}[tb]
\hspace*{-2cm}   
\includegraphics[width=12cm]{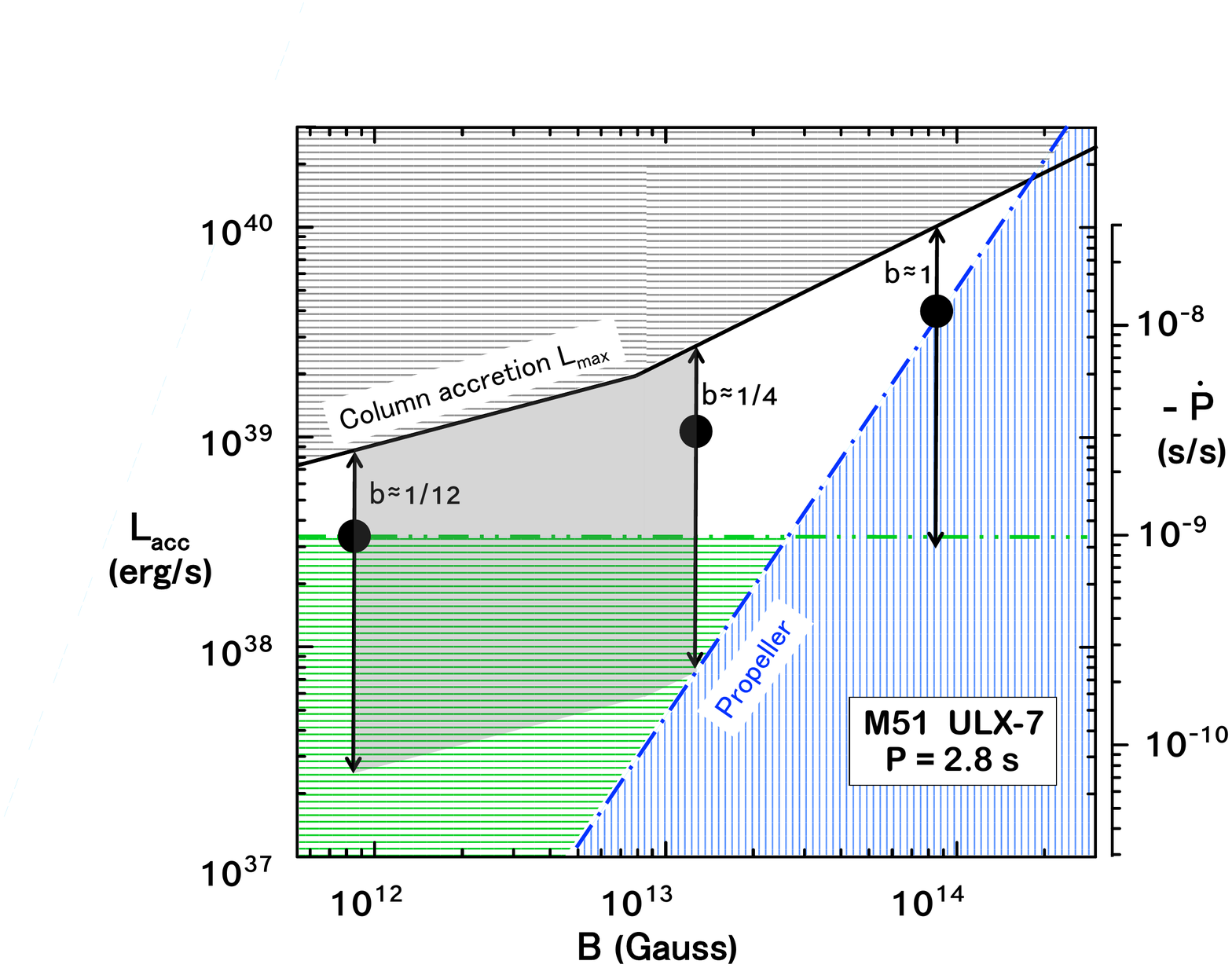}
\caption{
Accretion luminosity versus dipole magnetic field constraints for \src. 
The black solid line shows the maximum luminosity that magnetic column-accretion
onto the NS can attain \citep{mushtukov15}. 
The dot-dashed (blue) line marks the transition to the propeller regime, below which little (if any) accretion onto the NS surface can take place.
The secular $\dot{P}$ of \src, the double-dotted dashed (green) line (see the \mbox{Y-axis} scale on the right), is plotted here in correspondence with the minimum accretion luminosity that can give rise to it; below this line, the accretion rate would make the NS to spin-up at a lower rate than observed. 
Double-arrowed segments represent the factor of $\sim$30 luminosity variations observed from the source  under the assumption that they are due to accretion rate variations onto the NS surface ($L_{\mathrm{acc}}\propto \dot{M}$); the black circle in them shows the time-averaged luminosity $\langle L_{\mathrm{X}}\rangle$. 
Vertical shifts of the segments correspond to different values of the beaming factor $b = L_{\mathrm{acc}}/L_{\mathrm{iso}}$ and horizontal shifts to different values of the NS dipolar surface field.  
For accretion to take place unimpeded down to the lowest observed luminosities level, the bottom of the segment should be above the propeller line; for the maximum observed luminosity not to exceed the maximum luminosity of magnetic column accretion the top of the segment should be below the corresponding line. Moreover only segments positions for which the circle sits above the double-dotted dashed line are allowed (otherwise the time-averaged accretion rate would not be sufficient to secularly spin-up the NS at the observed rate). The leftmost and rightmost position of the segments for which all constraints are satisfied corresponds to beaming factors of $b\sim 1/12 $ and 1/4, respectively (see text). Therefore the gray-shaded area between $b\sim 1/12 $--1/4 and $B\sim 8\times10^{11}$--10$^{13}$\,G represent the allowed region for \src\ (see text).
If the assumption that the luminosity follows $\propto$\,$\dot{M}$ is relaxed, then reductions in luminosity may results at least in part from obscuration (e.g. disk precession), in which case the actual accretion luminosity range  would be smaller and confined close to the top of the double-arrowed segments. 
The allowed range, besides the gray shaded area, would then extend to the white region  up to $b\sim1$.}
\label{fig:propeller}
\end{figure}

Calculations by \citet{mushtukov15} show that a magnetically-funneled 
column accretion onto the NS poles can attain highly super-Eddington 
luminosities for very high 
magnetic fields (up to a few $10^3\,{L_{\mathrm{Edd}}}$
for \mbox{$B\sim5\times10^{15}$\,G}).
The maximum value that can be reached as a function of the NS
field at the magnetic poles is plotted as a solid line in Fig.\,\ref{fig:propeller}. 
In the same figure,  the dot-dashed line separates the region of the accretion regime (on the left) 
from that of the propeller regime (on the right) for a 2.8\,s-spinning 
accreting NS (our discussion here parallels the one in
\citealt{israel17}).

If \src\ emitted isotropically a maximum luminosity
$L_{\text{max,iso}}$\,$\sim$10$^{40}$\,\lum , then a NS 
magnetic field of about $\sim$\,$10^{14}$\,G would be 
required. The rightmost double-arrowed vertical segment in
Fig.\,\ref{fig:propeller} shows the factor of $\sim$30 luminosity range
observed from the source, with the black circle representing
the averaged value $\langle L_{\rm X}\rangle$ (inferred from the luminosity values reported in \citealt{earnshaw16} and in this work). 
It is apparent that for luminosities below $\langle L_{\rm X}\rangle$,
the source would straddle the transition to the propeller
regime: based on the discussion above, we deem 
this unlikely.

We consider the possibility that the emission of \src\ is collimated
over a fraction $b<1$ of the sky. The measured luminosity   
thus corresponds to the \textit{apparent} isotropic equivalent luminosity, 
and the accretion luminosity is thus
reduced according to  \mbox{$L_{\mathrm{acc}} = b\ L_{\mathrm{iso}}$}. 
By requiring that the propeller regime has not yet set in at the accretion
luminosity corresponding to the minimum detected isotropic
luminosity, {\it i.e.}  \mbox{$L_{\mathrm{min,acc}}=b\,L_{\mathrm{min,iso}}$}, 
and that  maximum accretion luminosity corresponding
the maximum isotropic luminosity \mbox{$L_{\mathrm{max,acc}}=b\,L_{\mathrm{max,iso}}$} 
is consistent with being produced by column accretion in accordance with \citet{mushtukov15}, we derive a maximum beaming factor of $b\sim 1/4$ and a maximum NS dipolar magnetic field of $\sim$\,$10^{13}$\,G (see Fig.\,8).  For any value of $b$ smaller 
than $\sim 1/4$ ({\it i.e.}, a more pronounced degree of beaming) 
a range of values of $B$\,$\leq$10$^{13}$\,G can be found such that 
the above two requirements are also verified. 
This may suggest that the beaming factor $b$ can attain very small 
values and the corresponding accretion luminosity be reduced at will. 
However an additional constraint comes from the observed spin-up rate
which must be sustained by a sufficiently high accretion rate. 
To work out the minimum accretion luminosity (and hence the accretion 
rate) required to give rise to a secular spin-up of  
$\dot{P} <-10^{-9}$\,s\,s$^{-1}$,
we consider that the highest specific angular momentum that can be
transferred to the NS is that of disk matter entering the NS magnetosphere 
at the corotation radius  $r_{\mathrm{co}}$. The resulting upper limit on  
the accretion torque translates into the condition $-\dot{P} < \dot{M}\ r_{\mathrm{co}}^2P/I$
(where \mbox{$I \sim 10^{45}$\,g cm$^2$} is the NS moment of inertia). 
This in turn gives \mbox{$\dot{M} > 3 \times 10^{18}$\,g s$^{-1}$} 
for the time-averaged accretion rate and, equivalently,  
$\langle L_{\rm acc}\rangle>3 \times 10^{38}$\,\lum, 
implying that $b> 1/12$ and $B>8\times10^{11}$\,G. 
Therefore we conclude that \src\ is likely a moderately beamed 
X-ray pulsar ($1/12 < b < 1/4$), accreting at up to 
$\sim$20 times the Eddington
rate and with a magnetic field between $\sim$$8 \times 10^{11}$ 
and $\sim$$10^{13}$\,G (see the grey-shaded area in Fig.\,\ref{fig:propeller}).
Note that we implicitly assumed that the
B-field is purely dipolar. Based on the above interpretation the properties of \src\ do not require (but cannot exclude) the presence of higher multipolar components dominating in the vicinity of the 
star surface; this is unlike the case of the PULX in NGC 5907, whose much larger spin up rate requires a higher $L_{acc}$ and thus powering by column accretion a high magnetic field (see \citealt{israel17}).


An alternative interpretation of PULX properties, based on models of disk-accreting 
black holes in ULXs, envisages that a large fraction of the super-Eddington 
mass inflow through the disk is ejected from within the radius where the disk becomes 
geometrically thick, with radiation escaping from a collimated funnel \citep{King09}. 
The accretion rate onto the NS surface 
is self-regulated close to the Eddington limit, such that magnetic-column accretion would work for
known PULX without ever resorting to high, magnetar-like B-fields \citep{king17}. 
We applied the above interpretation by using the prescription 
in \citet{king19} and the values of the maximum (isotropic) luminosity, spin, secular spin-up rate
measured  for \src. We derived values of the mass inflow rate through the magnetospheric 
boundary of about 12 times the Eddington rate and surface magnetic field of $\sim 1.4 \times 10^{12}$~G.
These values are not in tension  with the range of values inferred in our interpretation of \src,
despite clear differences in the modeling.

The soft X-ray spectral component of \src\ may originate 
in the accretion disk, as originally proposed to interpret 
the soft X-ray excess observed in a number of Galactic 
accreting pulsars in high mass X-ray binaries 
(e.g. the case of Her\,X-1, SMC\,X-1, LMC\,X-4, etc.; \citealt{hickox04}). Its luminosity, that we estimated from the best-fits of the average spectra of the bright states to be more than 30\% of the total, exceeds the energy release
of the disk itself by about 2 orders of magnitude; 
therefore, it is likely powered by reprocessing of the primary
central X-ray emission, the hard, pulsed component. 
Indeed, our {\sc diskbb} fits to the soft
component variations with a color correction (see above)give inner disk radii consistent with the corotation radius of 3300 km.
The equivalent blackbody radius of the hard component, being $\sim$5--10 times 
larger than the NS radius, may originate itself from reprocessing
by optically thick curtains of matter in the magnetic funnel 
that feeds the accretion column from higher altitudes. 
However, in order to intercept and reprocess $>$30\% of the accretion luminosity, the inner disk 
regions must subtend a large solid angle relative to the central source of primary radiation: 
it is not clear that such a puffed-up disk would maintain the same surface emissivity law of a standard disk, 
as implicit in the diskbb model. Substituting the {\sc diskbb} with a {\sc diskpbb} model in the fit of the average A, B and C spectra, 
we could obtain acceptable results with the emissivity index converging towards $p\sim0.5$. Hence, from our spectral analysis, we cannot exclude the existence of a thick disc in this system.

The observed luminosity swing --- encompassing also the disappearance (non detection)
of the harder component in the faint state, observation L --- together with the suggested 
$\approx$40\,d superorbital flux variations (from a \emph{Swift} monitoring; 
Brightman et al., in prep.) may be due to genuine variation of the mass accretion rate onto the NS 
(as assumed above) or result from partial obscuration of the X-ray emission relative to our line of sight, due, e.g. to a precessing accretion disk (see e.g. \citealt{middleton18}).
In the former case a recurrent disk instability or 
a modulation in the mass transfer rate from the companion induced by a third star in an eccentric 
$\sim$40\,d orbit, might cause the observed luminosity variations. 
In the latter case the nodal precession of 
a tilted disk may modulate the observed flux through partial obscuration (see e.g. the so-called slaved, disk model for the superorbital cycle of \mbox{Her\,X-1}, \citealt{roberts74}). The true luminosity swing of the source would be smaller and confined to the upper range close to $L_{\mathrm{iso,max}}$. In this case 
also the white region between $b\sim 1/4$ and $b\sim 1$ in Fig.\,\ref{fig:propeller} would be allowed and the NS magnetic field may be as high as $\sim$\,$10^{14}$\,G. 
In this interpretation, obscuration of the central X-ray source (hard spectral component) may be expected during the lowest flux intervals, as indeed observed in observation L.
Being driven by a changing inclination angle of the 
disk, the apparent luminosity variations of the soft component 
would be expected to scale approximately with the projected area along 
the line of sight, such that the inner disk radius 
from the diskbb fit should be $\propto$\,$L^{1/2}$ and its temperature constant. However in our fits the scaling with luminosity is approximately $R_{\mathrm{in}}\propto L^{-0.3}$ 
and the temperature dropped by a factor of $\sim$2 
in observation L. Nevertheless, it might be possible to reproduce these results through a suitably-shaped vertical profile of the precessing disk. 
Taken at face value the $TR_{\mathrm{in}}\propto L^{-0.3}$ dependence is consistent with the expected
dependence of the magnetospheric radius on the (accretion) luminosity, 
$r_{\mathrm{m}}\propto L^{-2/7}$. This may suggest that the luminosity swing 
of \src\ is driven by changes in the mass accretion rate onto the 
neutron star (rather than variations of viewing geometry), though in this scenario 
it is not clear how to account for the disappearance of the hard component in observation L. 

\subsection{The nature of the system}\label{nature:system}   

The timing parameters of \src\ firmly place the hosting binary system in the HMXB class, with a companion star of minimum mass $\sim 8\,M_\odot$. In the following we assume that the system undergoes Roche lobe overflow  (RLOF) with a donor close to the minimum estimated mass of 8$M_\odot$ and an orbital period of 2\,d. In this scenario, 
the radius and intrinsic luminosity of the donor at this stage are $\sim$7\,$R_\odot$ and $\sim$4000\,$L_\odot$, respectively (e.g. \citealt{eggleton83,demircan91}). However, owing to the large mass ratio \mbox{($q = M_{\mathrm{donor}}/M_{\mathrm{NS}} > 1$)}, the mass transfer may be unstable (e.g. \citealt{frank02}), if we approximately consider that the evolution proceeds on a thermal timescale $T_{\mathrm{th}}$, the duration of this phase is $\approx$50,000\,yr. The estimated average mass transfer rate is largely in excess of the Eddington limit ($\ga$20,000). Assuming a standard accretion efficiency  $\eta=(GM/c^2 R) \sim 10\%$ for a NS and negligible beaming, the observed X-ray luminosity ($\sim$\,$10^{40}$\,\lum) implies that only $\sim$3\% of the average transferred mass is in fact accreted onto the NS. Therefore, either the system is close to the propeller stage, and accretion is frequently interrupted/limited, or there are powerful outflows launched from the accretion disc, like those 
reported recently in a few ULXs (e.g. \citealt{pinto16,kosec18}).
For the total duration of the mass transfer phase the total mass deposited onto the NS is $\sim$0.1$M_\odot$. These numbers are consistent with the possibility that \src\ is an HMXB accreting above the Eddington limit. On the other hand, the evolution of a system like this can become dynamically unstable and rapidly lead to a common envelope phase.
Alternatively, a slightly different mechanism,  known as wind-RLOF, might be at work in \src, providing a stable mass transfer even for relatively large donor-to-accretor mass ratio, q$<$15, and with a similar average accreted mass ($\sim$3\%; \citealt{elmellah19}, see also \citealt{fragos15} for an alternative possible scenario).

\src\ dwells in the same region of the Corbet's diagram (spin versus orbital periods for accreting pulsars, see \citealt{enoto14} for a recent compilation) of the OB giant and supergiant HMXB systems in Roche lobe overflow in the Milky Way and the Magellanic Clouds.
These objects, Cen\,X-3 in our Galaxy, SMC\,X-1  and LMC\,X-4 (which we mentioned in connection with the superorbital flux and spectral variations in Sec.\,\ref{accretion:model} and for the pulsation drop-out in Sec.\,\ref{sec:signal}), are all rather bright NS pulsators, and the two sources in the Magellanic Clouds, remarkably, shine at, or slightly above, the Eddington luminosity (see e.g. \citealt{lutovinov13}, and \citealt{falanga15} for their orbital parameters and other characteristics of the systems). We note that the PULX M82\,X-2, with a spin period of about 1.4\,s and a orbital period of about 2.5\,d, lies in the same region of the diagram, though the optical counterpart is unknown.
The stellar classification of the donor in \src\ is unknown, but some of the candidate counterparts in \citet[][figure\,12]{earnshaw16} and in Sect.\,\ref{counterparts} have magnitude and colors consistent with OB supergiants (typical values are $M_V$ from --6 to --7, and $(B-V)$ from --0.1 to --0.4; e.g. \citealt{cox00}).


 However, as mentioned above, for Roche lobe overflow onto a NS with a donor close to the minimum estimated mass of 8 M$_\odot$, the system is probably in rapid and unstable evolution, and its optical emission is likely to be far from the expected photometric properties of a single isolated massive star. On the other hand, the optical properties of a wind-RLOF accreting system are not known in detail. In both cases, at the present stage it is difficult to make a meaningful comparison with the observed potential counterpart candidates.


\section{Conclusions}
\src\ is a newly identified PULX in a high mass X-ray binary ($M_\star\gtrsim8 M_\odot$) with an orbital period of 1.997 days and it is characterized by a spin signal with variable (on time scales of hours) properties. In particular, we note that the detection of such weak (few-percent PF) and variable signals in crowded fields can be challenging, probably limiting our chances to obtain a complete picture of the PULX demography with the current-generation X-ray missions. In this respect, the high throughput and good spatial/timing resolution of \emph{Athena} are expected to be  game changers in determining the NS incidence among ULXs.

Though our observational campaign allowed us to infer a relatively accurate orbital solution, other near future timing observations of \src\ are needed to further reduce the uncertainties in the orbital parameters and therefore improve the estimates of the spin parameters in the two epochs considered here. This might firmly establish if the propeller mechanism is at play in
this source or not. Besides, a better knowledge of the orbital parameters may turn the marginal detection of pulsation in the two archival observations into robust detections, enabling us to extend back by more than a decade the timing history of this PULX.

Our analysis suggest that a relatively ``standard'' dipolar magnetic field of 10$^{12}$--10$^{13}$\,G is sufficient to account for the observed luminosity, though we cannot exclude the presence of a stronger (up to $\sim 10^{14}$\,G) multipolar component close to the NS surface.

\acknowledgments
We thank Dr. Norbert Schartel for granting
DPS \xmm\ time.
The scientific results reported in this article are based on 
observations obtained with \xmm, an ESA science mission with 
instruments and contributions directly funded by ESA Member States 
and NASA.
GR acknowledges the support of high performance computing resources  
awarded by CINECA (MARCONI and GALILEO), under the ISCRA initiative 
and the INAF-CIENCA MoU; and also the computing centres of INAF -- Osservatorio Astronomico 
di Trieste and Osservatorio Astrofisico di Catania, under the 
coordination of the CHIPP project, for the availability of computing 
resources and support.
This work has also made use of observations made with the NASA/ESA 
\emph{Hubble Space Telescope}, and obtained from the Hubble Legacy 
Archive, which is a collaboration between the Space Telescope Science 
Institute (STScI/NASA), the Space Telescope European Coordinating Facility
(ST-ECF/ESA) and the Canadian Astronomy Data Centre (CADC/NRC/CSA).
This work has also made use of data from the ESA mission \emph{Gaia}, 
processed by the \emph{Gaia} Data Processing and Analysis Consortium
(DPAC).
We acknowledge funding in the framework of the project ULTraS
ASI--INAF contract N.\,2017-14-H.0, project ASI--INAF contract I/037/12/0, and 
PRIN grant 2017LJ39LM.
FB is funded by the European
Union’s Horizon 2020 research and innovation programme under
the Marie Sk\l odowska Curie grant agreement no. 664931.
FF and CP acknowledge support from ESA Research Fellowships.
This research was supported by high
performance computing  resources at New York University Abu Dhabi.
HPE acknowledges support under NASA contract NNG08FD60C.
TPR acknowledges support from STFC as part of the consolidated grant ST/K000861/1. We thank the anonymous referee for the helpful comments on the manuscript.

\bibliographystyle{aasjournal}
\bibliography{biblio}

\begin{thebibliography}{}
\expandafter\ifx\csname natexlab\endcsname\relax\def\natexlab#1{#1}\fi
\providecommand{\url}[1]{\href{#1}{#1}}
\providecommand{\dodoi}[1]{doi:~\href{http://doi.org/#1}{\nolinkurl{#1}}}
\providecommand{\doeprint}[1]{\href{http://ascl.net/#1}{\nolinkurl{http://ascl.net/#1}}}
\providecommand{\doarXiv}[1]{\href{https://arxiv.org/abs/#1}{\nolinkurl{https://arxiv.org/abs/#1}}}

\bibitem[{{Abolmasov} {et~al.}(2007){Abolmasov}, {Fabrika}, {Sholukhova}, \&
  {Afanasiev}}]{abolmasov07}
{Abolmasov}, P., {Fabrika}, S., {Sholukhova}, O., \& {Afanasiev}, V. 2007,
  Astrophysical Bulletin, 62, 36

\bibitem[{{Arnaud}(1996)}]{arnaud96}
{Arnaud}, K.~A. 1996, in Astronomical Society of the Pacific Conference Series,
  Vol. 101, Astronomical Data Analysis Software and Systems V, ed. G.~H.
  {Jacoby} \& J.~{Barnes} (ASP, San Francisco), 17--20

\bibitem[{{Bachetti} {et~al.}(2014){Bachetti}, {Harrison}, {Walton},
  {Grefenstette}, {Chakrabarty}, {F{\"u}rst}, {Barret}, {Beloborodov}, {Boggs},
  {Christensen}, {Craig}, {Fabian}, {Hailey}, {Hornschemeier}, {Kaspi},
  {Kulkarni}, {Maccarone}, {Miller}, {Rana}, {Stern}, {Tendulkar}, {Tomsick},
  {Webb}, \& {Zhang}}]{bachetti14}
{Bachetti}, M., {Harrison}, F.~A., {Walton}, D.~J., {et~al.} 2014, \nat, 514,
  202

\bibitem[{{Bai}(1992)}]{bai92}
{Bai}, T. 1992, \apj, 397, 584

\bibitem[{{Bertin} \& {Arnouts}(1996)}]{bertin96}
{Bertin}, E., \& {Arnouts}, S. 1996, \aaps, 117, 393

\bibitem[{{Blackburn}(1995)}]{blackburn95}
{Blackburn}, J.~K. 1995, in ASP Conf. Ser., (San Francisco, CA: ASP), Vol.~77,
  {Shaw}, R.~A. and {Payne}, H.~E. and {Hayes}, J.~J.~E., eds., Astronomical
  Data Analysis Software and Systems IV., 367

\bibitem[{{Brightman} {et~al.}(2018){Brightman}, {Balokovi{\'c}}, {Koss},
  {Alexander}, {Annuar}, {Earnshaw}, {Gandhi}, {Harrison}, {Hornschemeier},
  {Lehmer}, {Powell}, {Ptak}, {Rangelov}, {Roberts}, {Stern}, {Walton}, \&
  {Zezas}}]{brightman18}
{Brightman}, M., {Balokovi{\'c}}, M., {Koss}, M., {et~al.} 2018, \apj, 867, 110

\bibitem[{{Brumback} {et~al.}(2018){Brumback}, {Hickox}, {Bachetti},
  {Ballhausen}, {F{\"u}rst}, {Pike}, {Pottschmidt}, {Tomsick}, \&
  {Wilms}}]{brumback18}
{Brumback}, M.~C., {Hickox}, R.~C., {Bachetti}, M., {et~al.} 2018, \apjl, 861,
  L7

\bibitem[{{Campana} {et~al.}(2001){Campana}, {Gastaldello}, {Stella}, {Israel},
  {Colpi}, {Pizzolato}, {Orlandini}, \& {Dal Fiume}}]{campana01}
{Campana}, S., {Gastaldello}, F., {Stella}, L., {et~al.} 2001, \apj, 561, 924

\bibitem[{{Campana} \& {Stella}(2000)}]{campana00}
{Campana}, S., \& {Stella}, L. 2000, \apj, 541, 849

\bibitem[{{Campana} {et~al.}(2002){Campana}, {Stella}, {Israel}, {Moretti},
  {Parmar}, \& {Orlandini}}]{campana02}
{Campana}, S., {Stella}, L., {Israel}, G.~L., {et~al.} 2002, \apj, 580, 389

\bibitem[{{Campana} {et~al.}(2018){Campana}, {Stella}, {Mereghetti}, \& {de
  Martino}}]{campana18}
{Campana}, S., {Stella}, L., {Mereghetti}, S., \& {de Martino}, D. 2018, \aap,
  610

\bibitem[{{Carpano} {et~al.}(2018){Carpano}, {Haberl}, {Maitra}, \&
  {Vasilopoulos}}]{carpano18}
{Carpano}, S., {Haberl}, F., {Maitra}, C., \& {Vasilopoulos}, G. 2018, \mnras,
  476, L45

\bibitem[{{Chashkina} {et~al.}(2017){Chashkina}, {Abolmasov}, \&
  {Poutanen}}]{chashkina17}
{Chashkina}, A., {Abolmasov}, P., \& {Poutanen}, J. 2017, \mnras, 470, 2799

\bibitem[{{Colbert} \& {Mushotzky}(1999)}]{colbert99}
{Colbert}, E.~J.~M., \& {Mushotzky}, R.~F. 1999, \apj, 519, 89

\bibitem[{{Corbet}(1996)}]{corbet96}
{Corbet}, R.~H.~D. 1996, \apjl, 457, L31

\bibitem[{{Cowan} {et~al.}(2011){Cowan}, {Cranmer}, {Gross}, \&
  {Vitells}}]{cowan11}
{Cowan}, G., {Cranmer}, K., {Gross}, E., \& {Vitells}, O. 2011, European
  Physical Journal C, 71

\bibitem[{{Cox}(2000)}]{cox00}
{Cox}, A.~N. 2000, {Allen's astrophysical quantities} (4th ed.~Publisher: New
  York: AIP Press; Springer, 2000.~Editedy by Arthur N.~Cox.~ ISBN: 0387987460)

\bibitem[{{Dall'Osso} {et~al.}(2015){Dall'Osso}, {Perna}, \&
  {Stella}}]{dallosso15}
{Dall'Osso}, S., {Perna}, R., \& {Stella}, L. 2015, \mnras, 449, 2144

\bibitem[{{Demircan} \& {Kahraman}(1991)}]{demircan91}
{Demircan}, O., \& {Kahraman}, G. 1991, \apss, 181, 313

\bibitem[{{Dewangan} {et~al.}(2005){Dewangan}, {Griffiths}, {Choudhury},
  {Miyaji}, \& {Schurch}}]{dewangan05}
{Dewangan}, G.~C., {Griffiths}, R.~E., {Choudhury}, M., {Miyaji}, T., \&
  {Schurch}, N.~J. 2005, \apj, 635, 198

\bibitem[{{Earnshaw} {et~al.}(2016){Earnshaw}, {Roberts}, {Heil}, {Mezcua},
  {Walton}, {Done}, {Harrison}, {Lansbury}, {Middleton}, \&
  {Sutton}}]{earnshaw16}
{Earnshaw}, H.~M., {Roberts}, T.~P., {Heil}, L.~M., {et~al.} 2016, \mnras, 456,
  3840

\bibitem[{{Earnshaw} {et~al.}(2019){Earnshaw}, {Roberts}, {Middleton},
  {Walton}, \& {Mateos}}]{earnshaw18}
{Earnshaw}, H.~P., {Roberts}, T.~P., {Middleton}, M.~J., {Walton}, D.~J., \&
  {Mateos}, S. 2019, \mnras, 483, 5554

\bibitem[{{Eggleton}(1983)}]{eggleton83}
{Eggleton}, P.~P. 1983, \apj, 268, 368

\bibitem[{{Enoto} {et~al.}(2014){Enoto}, {Sasano}, {Yamada}, {Tamagawa},
  {Makishima}, {Pottschmidt}, {Marcu}, {Corbet}, {Fuerst}, \&
  {Wilms}}]{enoto14}
{Enoto}, T., {Sasano}, M., {Yamada}, S., {et~al.} 2014, \apj, 786, 127

\bibitem[{{Esposito} {et~al.}(2018){Esposito}, {Rea}, \& {Israel}}]{eri18}
{Esposito}, P., {Rea}, N., \& {Israel}, G.~L. 2018, in Timing Neutron Stars:
  Pulsations, Oscillations and Explosions, ed. T.~{Belloni}, M.~{Mendez}, \&
  C.~{Zhang}, ASSL, Springer, in press (preprint: astro-ph/1803.05716)

\bibitem[{{Fabbiano} {et~al.}(1992){Fabbiano}, {Kim}, \&
  {Trinchieri}}]{fabbiano92}
{Fabbiano}, G., {Kim}, D.-W., \& {Trinchieri}, G. 1992, \apjs, 80, 531

\bibitem[{{Falanga} {et~al.}(2015){Falanga}, {Bozzo}, {Lutovinov},
  {Bonnet-Bidaud}, {Fetisova}, \& {Puls}}]{falanga15}
{Falanga}, M., {Bozzo}, E., {Lutovinov}, A., {et~al.} 2015, \aap, 577

\bibitem[{{Feng} \& {Soria}(2011)}]{feng11}
{Feng}, H., \& {Soria}, R. 2011, \nar, 55, 166

\bibitem[{{Frank} {et~al.}(2002){Frank}, {King}, \& {Raine}}]{frank02}
{Frank}, J., {King}, A., \& {Raine}, D.~J. 2002, {Accretion Power in
  Astrophysics: Third Edition} (Cambridge: Cambridge University Press)

\bibitem[{{F{\"u}rst} {et~al.}(2016){F{\"u}rst}, {Walton}, {Harrison}, {Stern},
  {Barret}, {Brightman}, {Fabian}, {Grefenstette}, {Madsen}, {Middleton},
  {Miller}, {Pottschmidt}, {Ptak}, {Rana}, \& {Webb}}]{furst16}
{F{\"u}rst}, F., {Walton}, D.~J., {Harrison}, F.~A., {et~al.} 2016, \apjl, 831,
  L14

\bibitem[{{Gabriel} {et~al.}(2004){Gabriel}, {Denby}, {Fyfe}, {Hoar}, {Ibarra},
  {Ojero}, {Osborne}, {Saxton}, {Lammers}, \& {Vacanti}}]{gabriel04}
{Gabriel}, C., {Denby}, M., {Fyfe}, D.~J., {et~al.} 2004, in ASP Conf. Ser.
  (San Francisco, CA: ASP), Vol. 314, Astronomical Data Analysis Software and
  Systems (ADASS) XIII, ed. F.~{Ochsenbein}, M.~G. {Allen}, \& D.~{Egret}, 759

\bibitem[{{Gladstone} \& {Roberts}(2009)}]{gladstone09}
{Gladstone}, J.~C., \& {Roberts}, T.~P. 2009, \mnras, 397, 124

\bibitem[{{Hickox} {et~al.}(2004){Hickox}, {Narayan}, \& {Kallman}}]{hickox04}
{Hickox}, R.~C., {Narayan}, R., \& {Kallman}, T.~R. 2004, \apj, 614, 881

\bibitem[{{Hobbs} {et~al.}(2006){Hobbs}, {Edwards}, \& {Manchester}}]{hobbs06}
{Hobbs}, G.~B., {Edwards}, R.~T., \& {Manchester}, R.~N. 2006, \mnras, 369, 655

\bibitem[{{Illarionov} \& {Sunyaev}(1975)}]{illarionov75}
{Illarionov}, A.~F., \& {Sunyaev}, R.~A. 1975, \aap, 39, 185

\bibitem[{{Israel} \& {Stella}(1996)}]{israel96}
{Israel}, G.~L., \& {Stella}, L. 1996, \apj, 468, 369

\bibitem[{{Israel} {et~al.}(2017{\natexlab{a}}){Israel}, {Belfiore}, {Stella},
  {Esposito}, {Casella}, {De Luca}, {Marelli}, {Papitto}, {Perri}, {Puccetti},
  {Castillo}, {Salvetti}, {Tiengo}, {Zampieri}, {D'Agostino}, {Greiner},
  {Haberl}, {Novara}, {Salvaterra}, {Turolla}, {Watson}, {Wilms}, \&
  {Wolter}}]{israel17}
{Israel}, G.~L., {Belfiore}, A., {Stella}, L., {et~al.} 2017{\natexlab{a}},
  Science, 355, 817

\bibitem[{{Israel} {et~al.}(2017{\natexlab{b}}){Israel}, {Papitto}, {Esposito},
  {Stella}, {Zampieri}, {Belfiore}, {Rodr{\'{\i}}guez Castillo}, {De Luca},
  {Tiengo}, {Haberl}, {Greiner}, {Salvaterra}, {Sandrelli}, \&
  {Lisini}}]{ipe17}
{Israel}, G.~L., {Papitto}, A., {Esposito}, P., {et~al.} 2017{\natexlab{b}},
  \mnras, 466, L48

\bibitem[{{Kaaret} {et~al.}(2017){Kaaret}, {Feng}, \& {Roberts}}]{kaaret17}
{Kaaret}, P., {Feng}, H., \& {Roberts}, T.~P. 2017, \araa, 55, 303

\bibitem[{{King} \& {Lasota}(2016)}]{king16}
{King}, A., \& {Lasota}, J.-P. 2016, \mnras, 458, L10

\bibitem[{{King} {et~al.}(2017){King}, {Lasota}, \& {Klu{\'z}niak}}]{king17}
{King}, A., {Lasota}, J.-P., \& {Klu{\'z}niak}, W. 2017, \mnras, 468, L59

\bibitem[{{Koliopanos} {et~al.}(2017){Koliopanos}, {Vasilopoulos}, {Godet},
  {Bachetti}, {Webb}, \& {Barret}}]{koliopanos17}
{Koliopanos}, F., {Vasilopoulos}, G., {Godet}, O., {et~al.} 2017, \aap, 608,
  A47

\bibitem[{{Kosec} {et~al.}(2018){Kosec}, {Pinto}, {Walton}, {Fabian},
  {Bachetti}, {Brightman}, {F{\"u}rst}, \& {Grefenstette}}]{kosec18}
{Kosec}, P., {Pinto}, C., {Walton}, D.~J., {et~al.} 2018, \mnras, 479, 3978

\bibitem[{{Kuntz} {et~al.}(2016){Kuntz}, {Long}, \& {Kilgard}}]{kuntz16}
{Kuntz}, K.~D., {Long}, K.~S., \& {Kilgard}, R.~E. 2016, \apj, 827, 46

\bibitem[{{Lange} {et~al.}(2001){Lange}, {Camilo}, {Wex}, {Kramer}, {Backer},
  {Lyne}, \& {Doroshenko}}]{lange01}
{Lange}, C., {Camilo}, F., {Wex}, N., {et~al.} 2001, \mnras, 326, 274

\bibitem[{{Leahy} {et~al.}(1983){Leahy}, {Darbro}, {Elsner}, {Weisskopf},
  {Kahn}, {Sutherland}, \& {Grindlay}}]{leahy83}
{Leahy}, D.~A., {Darbro}, W., {Elsner}, R.~F., {et~al.} 1983, \apj, 266, 160

\bibitem[{{Liu} {et~al.}(2002){Liu}, {Bregman}, {Irwin}, \& {Seitzer}}]{liu02}
{Liu}, J.-F., {Bregman}, J.~N., {Irwin}, J., \& {Seitzer}, P. 2002, \apjl, 581,
  L93

\bibitem[{{Liu} \& {Mirabel}(2005)}]{liu05}
{Liu}, Q.~Z., \& {Mirabel}, I.~F. 2005, \aap, 429, 1125

\bibitem[{{Long} \& {van Speybroeck}(1983)}]{long83}
{Long}, K.~S., \& {van Speybroeck}, L.~P. 1983, in Accretion-Driven Stellar
  X-ray Sources, Cambridge University Press, Cambridge., ed. W.~H.~G. {Lewin}
  \& E.~P.~J. {van den Heuvel}, 141

\bibitem[{{Lutovinov} {et~al.}(2013){Lutovinov}, {Revnivtsev}, {Tsygankov}, \&
  {Krivonos}}]{lutovinov13}
{Lutovinov}, A.~A., {Revnivtsev}, M.~G., {Tsygankov}, S.~S., \& {Krivonos},
  R.~A. 2013, \mnras, 431, 327

\bibitem[{{McQuinn} {et~al.}(2016){McQuinn}, {Skillman}, {Dolphin}, {Berg}, \&
  {Kennicutt}}]{mcquinn16}
{McQuinn}, K.~B.~W., {Skillman}, E.~D., {Dolphin}, A.~E., {Berg}, D., \&
  {Kennicutt}, R. 2016, \apj, 826, 21

\bibitem[{{Middleton} {et~al.}(2015){Middleton}, {Heil}, {Pintore}, {Walton},
  \& {Roberts}}]{middleton15}
{Middleton}, M.~J., {Heil}, L., {Pintore}, F., {Walton}, D.~J., \& {Roberts},
  T.~P. 2015, \mnras, 447, 3243

\bibitem[{{Middleton} \& {King}(2017{\natexlab{a}})}]{middleton17}
{Middleton}, M.~J., \& {King}, A. 2017{\natexlab{a}}, \mnras, 470, L69

\bibitem[{{Middleton} \& {King}(2017{\natexlab{b}})}]{middleton17e}
---. 2017{\natexlab{b}}, \mnras, 471, L71

\bibitem[{{Middleton} {et~al.}(2018){Middleton}, {Fragile}, {Bachetti},
  {Brightman}, {Jiang}, {Ho}, {Roberts}, {Ingram}, {Dauser}, {Pinto}, {Walton},
  {Fuerst}, {Fabian}, \& {Gehrels}}]{middleton18}
{Middleton}, M.~J., {Fragile}, P.~C., {Bachetti}, M., {et~al.} 2018, \mnras,
  475, 154

\bibitem[{{Miller} {et~al.}(2003){Miller}, {Wijnands}, {M{\'e}ndez},
  {Kendziorra}, {Tiengo}, {van der Klis}, {Chakrabarty}, {Gaensler}, \&
  {Lewin}}]{miller03}
{Miller}, J.~M., {Wijnands}, R., {M{\'e}ndez}, M., {et~al.} 2003, \apjl, 583,
  L99

\bibitem[{{Mitsuda} {et~al.}(1984){Mitsuda}, {Inoue}, {Koyama}, {Makishima},
  {Matsuoka}, {Ogawara}, {Suzuki}, {Tanaka}, {Shibazaki}, \&
  {Hirano}}]{mitsuda84}
{Mitsuda}, K., {Inoue}, H., {Koyama}, K., {et~al.} 1984, \pasj, 36, 741

\bibitem[{Murphy \& Vaart(2000)}]{murphy00}
Murphy, S.~A., \& Vaart, A. W. V.~D. 2000, Journal of the American Statistical
  Association, 95, 449

\bibitem[{{Mushtukov} {et~al.}(2015){Mushtukov}, {Suleimanov}, {Tsygankov}, \&
  {Poutanen}}]{mushtukov15}
{Mushtukov}, A.~A., {Suleimanov}, V.~F., {Tsygankov}, S.~S., \& {Poutanen}, J.
  2015, \mnras, 454, 2539

\bibitem[{{Palumbo} {et~al.}(1985){Palumbo}, {Fabbiano}, {Fransson}, \&
  {Trinchieri}}]{palumbo85}
{Palumbo}, G.~G.~C., {Fabbiano}, G., {Fransson}, C., \& {Trinchieri}, G. 1985,
  \apj, 298, 259

\bibitem[{{Pinto} {et~al.}(2016){Pinto}, {Middleton}, \& {Fabian}}]{pinto16}
{Pinto}, C., {Middleton}, M.~J., \& {Fabian}, A.~C. 2016, \nat, 533, 64

\bibitem[{{Pintore} {et~al.}(2015){Pintore}, {Esposito}, {Zampieri}, {Motta},
  \& {Wolter}}]{pintore15}
{Pintore}, F., {Esposito}, P., {Zampieri}, L., {Motta}, S., \& {Wolter}, A.
  2015, \mnras, 448, 1153

\bibitem[{{Pintore} {et~al.}(2017){Pintore}, {Mereghetti}, {Tiengo},
  {Vianello}, {Costantini}, \& {Esposito}}]{Pintore17}
{Pintore}, F., {Mereghetti}, S., {Tiengo}, A., {et~al.} 2017, \mnras, 467, 3467

\bibitem[{{Pintore} {et~al.}(2014){Pintore}, {Zampieri}, {Wolter}, \&
  {Belloni}}]{pintore14}
{Pintore}, F., {Zampieri}, L., {Wolter}, A., \& {Belloni}, T. 2014, \mnras,
  439, 3461

\bibitem[{{Roberts}(2007)}]{roberts07}
{Roberts}, T.~P. 2007, \apss, 311, 203

\bibitem[{{Roberts} \& {Warwick}(2000)}]{roberts00}
{Roberts}, T.~P., \& {Warwick}, R.~S. 2000, \mnras, 315, 98

\bibitem[{{Roberts}(1974)}]{roberts74}
{Roberts}, W.~J. 1974, \apj, 187, 575

\bibitem[{{Sathyaprakash} {et~al.}(2019){Sathyaprakash}, {Roberts}, {Walton},
  {Fuerst}, {Bachetti}, {Pinto}, {Alston}, {Earnshaw}, {Fabian}, \&
  {Middleton}}]{sath19}
{Sathyaprakash}, R., {Roberts}, T.~P., {Walton}, D.~J., {et~al.} 2019, arXiv
  e-prints, arXiv:1906.00640

\bibitem[{{Stauffer}(1982)}]{stauffer82}
{Stauffer}, J.~R. 1982, \apj, 262, 66

\bibitem[{{Stella} {et~al.}(1986){Stella}, {White}, \& {Rosner}}]{stella86}
{Stella}, L., {White}, N.~E., \& {Rosner}, R. 1986, \apj, 308, 669

\bibitem[{{Stobbart} {et~al.}(2006){Stobbart}, {Roberts}, \&
  {Wilms}}]{stobbart06}
{Stobbart}, A.-M., {Roberts}, T.~P., \& {Wilms}, J. 2006, \mnras, 368, 397

\bibitem[{{Str{\"u}der} {et~al.}(2001){Str{\"u}der}, {Briel}, {Dennerl},
  {Hartmann}, {Kendziorra}, {Meidinger}, {Pfeffermann}, {Reppin}, {Aschenbach},
  {Bornemann}, {Br{\"a}uninger}, {Burkert}, {Elender}, {Freyberg}, {Haberl},
  {Hartner}, {Heuschmann}, {Hippmann}, {Kastelic}, {Kemmer}, {Kettenring},
  {Kink}, {Krause}, {M{\"u}ller}, {Oppitz}, {Pietsch}, {Popp}, {Predehl},
  {Read}, {Stephan}, {St{\"o}tter}, {Tr{\"u}mper}, {Holl}, {Kemmer}, {Soltau},
  {St{\"o}tter}, {Weber}, {Weichert}, {von Zanthier}, {Carathanassis}, {Lutz},
  {Richter}, {Solc}, {B{\"o}ttcher}, {Kuster}, {Staubert}, {Abbey}, {Holland},
  {Turner}, {Balasini}, {Bignami}, {La Palombara}, {Villa}, {Buttler},
  {Gianini}, {Lain{\'e}}, {Lumb}, \& {Dhez}}]{struder01}
{Str{\"u}der}, L., {Briel}, U., {Dennerl}, K., {et~al.} 2001, \aap, 365, L18

\bibitem[{{Sutton} {et~al.}(2013){Sutton}, {Roberts}, \&
  {Middleton}}]{sutton13}
{Sutton}, A.~D., {Roberts}, T.~P., \& {Middleton}, M.~J. 2013, \mnras, 435,
  1758

\bibitem[{{Sutton} {et~al.}(2012){Sutton}, {Roberts}, {Walton}, {Gladstone}, \&
  {Scott}}]{sutton12}
{Sutton}, A.~D., {Roberts}, T.~P., {Walton}, D.~J., {Gladstone}, J.~C., \&
  {Scott}, A.~E. 2012, \mnras, 423, 1154

\bibitem[{{Terashima} \& {Wilson}(2004)}]{terashima04}
{Terashima}, Y., \& {Wilson}, A.~S. 2004, \apj, 601, 735

\bibitem[{{Trudolyubov}(2008)}]{trudolyubov08}
{Trudolyubov}, S.~P. 2008, \mnras, 387, L36

\bibitem[{{Trudolyubov} {et~al.}(2007){Trudolyubov}, {Priedhorsky}, \&
  {C{\'o}rdova}}]{trudolyubov07}
{Trudolyubov}, S.~P., {Priedhorsky}, W.~C., \& {C{\'o}rdova}, F.~A. 2007, \apj,
  663, 487

\bibitem[{{Tsygankov} {et~al.}(2016){Tsygankov}, {Mushtukov}, {Suleimanov}, \&
  {Poutanen}}]{tsygankov16}
{Tsygankov}, S.~S., {Mushtukov}, A.~A., {Suleimanov}, V.~F., \& {Poutanen}, J.
  2016, \mnras, 457, 1101

\bibitem[{{Turner} {et~al.}(2001){Turner}, {Abbey}, {Arnaud}, {Balasini},
  {Barbera}, {Belsole}, {Bennie}, {Bernard}, {Bignami}, {Boer}, {Briel},
  {Butler}, {Cara}, {Chabaud}, {Cole}, {Collura}, {Conte}, {Cros}, {Denby},
  {Dhez}, {Di Coco}, {Dowson}, {Ferrando}, {Ghizzardi}, {Gianotti}, {Goodall},
  {Gretton}, {Griffiths}, {Hainaut}, {Hochedez}, {Holland}, {Jourdain},
  {Kendziorra}, {Lagostina}, {Laine}, {La Palombara}, {Lortholary}, {Lumb},
  {Marty}, {Molendi}, {Pigot}, {Poindron}, {Pounds}, {Reeves}, {Reppin},
  {Rothenflug}, {Salvetat}, {Sauvageot}, {Schmitt}, {Sembay}, {Short},
  {Spragg}, {Stephen}, {Str{\"u}der}, {Tiengo}, {Trifoglio}, {Tr{\"u}mper},
  {Vercellone}, {Vigroux}, {Villa}, {Ward}, {Whitehead}, \& {Zonca}}]{turner01}
{Turner}, M.~J.~L., {Abbey}, A., {Arnaud}, M., {et~al.} 2001, \aap, 365, L27

\bibitem[{{Turolla} {et~al.}(2015){Turolla}, {Zane}, \& {Watts}}]{turolla15}
{Turolla}, R., {Zane}, S., \& {Watts}, A.~L. 2015, Reports on Progress in
  Physics, 78, 116901

\bibitem[{{Verner} {et~al.}(1996){Verner}, {Ferland}, {Korista}, \&
  {Yakovlev}}]{verner96}
{Verner}, D.~A., {Ferland}, G.~J., {Korista}, K.~T., \& {Yakovlev}, D.~G. 1996,
  \apj, 465, 487

\bibitem[{{Walton} {et~al.}(2018){Walton}, {F{\"u}rst}, {Heida}, {Harrison},
  {Barret}, {Stern}, {Bachetti}, {Brightman}, {Fabian}, \&
  {Middleton}}]{walton18b}
{Walton}, D.~J., {F{\"u}rst}, F., {Heida}, M., {et~al.} 2018, \apj, 856

\bibitem[{{Wiktorowicz} {et~al.}(2017){Wiktorowicz}, {Sobolewska}, {Lasota}, \&
  {Belczynski}}]{wiktorowicz17}
{Wiktorowicz}, G., {Sobolewska}, M., {Lasota}, J.-P., \& {Belczynski}, K. 2017,
  \apj, 846, 17

\bibitem[{{Wilms} {et~al.}(2000){Wilms}, {Allen}, \& {McCray}}]{wilms00}
{Wilms}, J., {Allen}, A., \& {McCray}, R. 2000, \apj, 542, 914

\bibitem[{{Zampieri} \& {Roberts}(2009)}]{zampieri09}
{Zampieri}, L., \& {Roberts}, T.~P. 2009, \mnras, 400, 677

\end{thebibliography}

\end{document}